\documentstyle[multicol,aps,prl,eqsecnum,floats,epsf]{revtex}
\begin{document}
\renewcommand{\thesection}{\arabic{section}}
\draft
\title{{\tenrm\hfill }\\
Instanton for random advection}
\author{
Michael Chertkov\cite{Prin}}
\address{
Department of Physics of Complex Systems, Weizmann Institute of Science,
Rehovot 76100, Israel}
\date{\today}
\maketitle

\begin{abstract}
A path integral over trajectories of $2n$ fluid particles is identified
with a $2n$-th order correlation function of a passive scalar
convected by $d$-dimensional short-correlated
multi-scale incompressible random velocity flow.
Strong intermittency of the scalar is described 
by means of an instanton calculus (saddle point $+$ fluctuations
about it) in the path integral at $n\gg d$.
Anomalous scaling exponent of the $2n$-th scalar's structural function
is found analytically. 
\end{abstract}

\section{Introduction}

The problem of scaling behavior in Kraichnan's model of a white-advected
passive scalar \cite{68Kra-a}, attracts a great deal of attention
\cite{94Kra,95KYC,95CFKLb,95GK,95CF,96BGK}.
In the wide range of scales, called the convective interval,
structural functions of the Lagrangian tracer $\theta$,
passively advected by $d$-dimensional
short-correlated in time multiscale incompressible flow, 
possess a scaling behavior. The anomalous scaling exponent $\zeta_{2n}$ of the
$2n$-th order structural function,
$\langle(\theta_1-\theta_2)^{2n}\rangle\propto r^{\zeta_{2n}}$,
has been calculated in the following cases:
i) large space dimensionality $\zeta_2 d\gg (2-\zeta_2)n$, 
for $n=2$ in \cite{95CFKLb},
and generally for all allowed $n$ in \cite{95CF}, $\zeta_{2n}\to 
2n(n-1)(2-\zeta_2)/d$;
ii) almost smooth scalar field $2-\zeta_2\ll 1$, $d>2$ for $n=2$ in \cite{95GK}
and generally for $n(2-\zeta_2)\ll d$, $d>n$ in \cite{96BGK},
$\zeta_{2n}\to 2n(n-1)(2-\zeta_2)/(d+2) $.
The perturbation methods yield the scaling exponents in the limits 
where the respective
bare approximations are strictly Gaussian and
the anomalous corrections are small.

Instanton (steepest descent) formalism 
after perturbation expansion is the second 
quantitative method that could be applied to
a general statistical problem. The method works when a large parameter
makes some very special rare configuration to have an exponentially
large weight. Such a large parameter may be a high order $n$ of correlation function
$\langle\varphi^n\rangle$ of fluctuating field $\varphi$. 
The bare instanton approximation is obviously
strongly Non-Gaussian.
The idea was originally introduced and
successfully applied in field theory \cite{77Lip}
almost twenty years ago, but introduced to the turbulence theory
only very recently. 
An instanton calculus in a Lagrangian path integral 
was used to find an exponential tail of the scalar's probability
distribution function 
(reflected intermittent, non-Gaussian behavior of higher moments)
in the case $\zeta_2=0$ of linear velocity profile \cite{94SS}
(later on it was shown that the limit
turns out to be solvable exactly
\cite{95CFKLa,94CGK,95BCKL}).
A general method for finding the non-Gaussian tails of pdf
for solutions of a stochastic differential equation,
such as the convection equation for a passive scalar, random driven Navier-Stokes
etc., was formulated in \cite{95FKLM}. The initial idea of the method is to look for
a saddle point configuration in the path integral for the generating functional
introduced in \cite{76Dom,76Jan}. 
The extremum of the effective action is given by a coupled field-force
configuration (instanton), varying in space and time.
The method was applied recently to Burgers' turbulence
\cite{95GM,96BFKL}.
Generally, it is very difficult to solve the coupled 
(field-force) instanton equations.

In the present paper we generalize the idea of \cite{94SS} 
for the case of a nonlinear velocity profile ($\zeta_2\!>\!0$).
 $\zeta_{2n}$ is calculated
for $n$ being the largest number in the problem.
The method is based on a
very special feature of the problem \cite{Kra,94SS}:
there exists a closed differential
equation connecting $2n$-th and $(2n\!-\!2)$-th simultaneous correlation
functions of the scalar.
The $2n$-th correlator 
is expressed via the convolution of the resolvent of the eddy-diffusivity operator
with a source function constructed from the $(2n\!-\!2)$-th correlation function.
To prepare a path-integral for the instanton
calculus, we perform an explicit map
of the original problem 
of  calculation of the $2n$-th order correlation
function to the problem of calculation of a matrix element in an auxiliary
$2n$-particle Quantum mechanics.
The resolvent of the eddy-diffusivity operator is 
expressed in the method 
via the path integral over trajectories of $2n$ fluid
particles moving from an initial geometry (at which we are aimed to
describe scalar's correlations) with a characteristic scale $r$
to a final large-scale ($\sim L$) geometry.
The tensor of eddy diffusivity plays the role of tensor of inverse mass
for the particles from the associated Quantum mechanics.
The tensor depends explicitly on relative distances between the particles. 

It is the large number  of particles that makes the auxiliary 
Quantum mechanics almost ``classical'' (``semi-classical'').
A ``classical'' $2n$-particle configuration 
is the desirable rare event that describes both the intermittency
of $2n$-th moments of scalar differences and 
intermittency of the $n$-th moment of the dissipation field 
$\varepsilon=\kappa(\nabla\theta)^2$
(it is proven in \cite{95CF} that they are related to each other,
if scale-invariance of the structural function and of
the correlator of the dissipation field is valid:
$\langle\varepsilon^n\rangle\sim[L/r_d]^{\Delta_{2n}}$,
$\Delta_{2n}=n\zeta_2-\zeta_{2n}$;
$\kappa$ and $r_d$ are the diffusion coefficient and scale, respectively).
Calculation of the ``classical'' (saddle-point) 
contribution into $\langle\varepsilon^n\rangle$ gives the scale-invariant 
answer: $\Delta_{2n}^{cl}\to n\zeta_2$ at $n\to\infty$. 
The ``classical'' anomaly shows the highest
level of intermittency possible: $\zeta_\infty^{cl}=0$.
Therewith exist a wide set of ``classical'' trajectories 
(realizing themselves separately,
for different initial displacements of the points and
different forms of the scalar's source) responsible
for the ``classical'' answer. 
To extract an optimal trajectory 
(that gives the lowest possible contribution of
fluctuations) from the set of the ``classical'' ones
and thus to get a finite asymptotic for $\zeta_{2n}$, 
we must account for the fluctuations about the saddle-points.

It is shown that, at $0\!<\!\zeta_2\!<\!1$, 
the optimal trajectory is defined as a relative dispersion
of two groups (drops) of
particles: there is one distance 
(separation between the drops) being stretched while all
the other distances (sizes of the drops) are being contracted dynamically.
They are Gaussian fluctuations about the optimal trajectory
that should give the true value of $\zeta_{2n}$.
Accounting for the relative longitudinal 
(along the ``classical'' stretching direction)
Gaussian fluctuations of the drops, gives the dominant contribution into
the $n$-independent asymptotics for $\zeta_{2n}$ at $n\to\infty$.
The exponent is finite, it 
grows linearly with $d$ and decreases monotonically with an increase
in $\zeta_2$. The finite limit for $\zeta_{2n}$ at $\zeta_2\to 0$
along with the nonanomalous answer $\zeta_{2n}=0$ for the strictly
``logarithmic'' limit $\zeta_2=0$ \cite{94SS,95CFKLa} show together
a discontinuity of $\zeta_{2n}$ at $\zeta_2=0$.
There exists a simple physical picture that explains the origin
of this discontinuity.
In the first case of a linear velocity profile,
distances between all the fluid particles
are stretched homotopically: there is no way
for two groups of particles to
diverge from each other and to keep the inner group distances
contracted (or even intact) simultaneously. 
Vice versa, in the case of finite $\zeta_2$
(yet $\zeta_2$ should be smaller than unity),
the two-point trajectory, with 
the sizes of the drops of particles being contracted dynamically
whereas the distance between the drops being increased, is allowed.

Most of Non-Gaussian fluctuations 
about the saddle-point configuration can be dropped
in comparison with the Gaussian ones if $n\gg \mbox{Pe},d$
(we should only worry about the explicit calculation of the
Non-Gaussian fluctuations corresponding to a soft rotation mode).
This method is not applicable for $\mbox{Pe}$ being of the order of
(the more so as being larger than) $n$.
However, making use of an overall observation (concerning
the linearity of the problem and the scale invariance feature
of different terms entered in the correlation functions) one can 
extend the anomalous result (but not the method used for its derivation) to the
limit $\mbox{Pe},n\gg d$ too. 

The two-point configuration is not relevant at $1\!<\!\zeta_2\!<\!2$
(repulsion of particles inside of a drop
is no longer weak to make the configuration stable dynamically).
Only trajectories with many ($\sim n$) distances being diverged
should be taken into account.
However, calculation of fluctuations about
such trajectories shows a strong renormalization of the saddle-point answer:
it is a product of $\sim n$ algebraic terms 
(each responsible for fluctuation of a distance)
that makes the contribution of fluctuations 
competitive with (or even large than) the ``classical'' value.
The resulting contribution into $\langle\varepsilon^n\rangle$
is negligible in comparison with the normal scaling term, that always exists. 
To conclude, the instanton calculus is not an appropriate tool in this case.

The material in this paper is organized as follows.
In Section II, after a detailed and formal definition of the problem
we introduce path integral representation for
the $2n$-th order correlation function of the passive scalar.
We present the path integral for $\langle\varepsilon^n\rangle$ too.
It completes preparation for delivering an instanton (steepest-descent)
formalism for calculation of $\langle\varepsilon^n\rangle$ at $n\gg d$
in the two forthcoming Sections. Saddle-point equations 
are derived and studied in Section III.
The contributions of different saddle-points 
into $\langle\varepsilon^n\rangle$ are calculated 
(Appendix A) and compared with each other in Subsection IIIB.
To improve the saddle-point calculations and to extract among the
saddle-points an optimal one we study Gaussian fluctuations about
the saddle-points in Section IV and Appendises B,C.
The anomalous exponent $\zeta_{2n}$ for the optimal saddle-point is
calculated there in Section IV.
In the concluding Section V we discuss the results
from the points of view of 
criteria of their applicability, restrictions imposed,
possible generalizations, and comparison with other results and methods.

\section{Formulation of the problem}
 
The advection of passive scalar is governed by the following equation
\begin{equation}
(\partial_t+v_\alpha\nabla_\alpha-\kappa\triangle)\theta=f,
\quad \nabla_\alpha v_\alpha=0,
\label{e1}
\end{equation}
where $f(t;{\bf r})$ is the external source,
${\bf v}(t;{\bf r})$ is the advecting $d$-dimensional velocity and $\kappa$ is
the diffusion coefficient. $f(t;{\bf r})$ and ${\bf v}(t;{\bf r})$
are independent random functions of $t$ and ${\bf r}$, 
both Gaussian and $\delta$-correlated in time.
The source is spatially correlated on a scale of the pumping $L$, i.e.
the pair correlation function 
$\langle f(t_1;{\bf r}_1)f(t_2;{\bf r}_2)\rangle=\delta(t_1-t_2)\chi(r_{12})$
as a function of its argument decays on the scale $L$.
The value of $\chi(0)=P$ is the production rate of $\theta^2$.
The eddy diffusivity tensor ${\cal K}^{\alpha\beta}$, that describes
the Gaussian velocity correlations,
\begin{eqnarray}&&
\langle v^\alpha(t_1;{\bf r}_1) v^\beta(t_2;{\bf r}_2)\rangle=
\delta(t_1-t_2)\left[V_0\delta^{\alpha\beta}-
{\cal K}^{\alpha\beta}({\bf r}_1-{\bf r}_2)\right],\
\label{2}\\&&
{\cal K}^{\alpha\beta}({\bf r})=\frac{D}{(2-\gamma)r^\gamma}\left[
(d+1-\gamma)\delta^{\alpha\beta} r^2-(2-\gamma)r^\alpha r^\beta\right],
\label{2a}
\end{eqnarray}
depends on two parameters: $D$ that defines
the level of turbulence and $\gamma$, $0<\gamma<2$, that measures a
degree of nonsmoothness of the velocity field.

Averaging (\ref{e1}) over the statistics of ${\bf u}(t;{\bf r})$ and 
$f(t;{\bf r})$, one gets the closed equation for the simultaneous 
correlation functions of the scalar 
$F_{1\cdot 2n}=\langle\theta({\bf r}_1)\cdots\theta({\bf r}_n)\rangle$ \cite{94SS}:
\begin{eqnarray}&&
-\hat{\cal L}_{2n} F_{1\cdots 2n}=\chi_{1\cdots 2n},\label{1a}\\
&&\chi_{1\cdots 2n}=\chi_{12}F_{3\cdots 2n}+\mbox{permutations},
\label{1b}\\
&& \hat{\cal L}_n=-\sum_{i\neq j}^n
 {\cal K}^{\alpha\beta}({\bf r}_i-{\bf r}_j)\nabla_i^\alpha\nabla_j^\beta+
\kappa\sum_i^n\triangle_i .\label{1c}
\end{eqnarray}
The dependence of the
source function $\chi_{2n}(r_{ij}\sim L)$ on $L$ at $r_{ij}\lesssim L$
is estimated as $\sim L^{(n-1)\gamma}$;
upscales from $L$ the function decays algebraically fast. 
It is the major information about
$\chi_{2n}$ required for further consideration.

The equation (\ref{1b}) for the pair correlation function ($n=1$) was
solved explicitly \cite{68Kra-a}. The pair correlator in the convective interval,
$r_d\ll r_{12}\ll L$, where $r_d^{2-\gamma}=2(2-\gamma)\kappa/[D(d-1)]$, 
gets the following form
\begin{equation}
\langle\theta_1\theta_2\rangle=P\frac{2-\gamma}{\gamma(d-1)D}\left(
\frac{L^\gamma}{d-\gamma}-\frac{r^\gamma}{d}\right).
\label{pc}
\end{equation}
Thus, the pair structural function is shown to have a simple scaling behavior
in the convective interval,
$\langle(\theta_1-\theta_2)^2\rangle\sim P r^{\gamma}/D$,
$\zeta_2=\gamma$, to provide the constancy of the flux of $\theta^2$ there.
The scaling exponent of $\hat{\cal L}$ is $-\gamma$, 
the function $\chi_{12}$ does not depend on $r_{12}$ deep 
inside the convective interval so that its exponent is $0$,
the solution of (\ref{pc}) thus may be presented in the form $F_{forc}+{\cal Z}$,
where we separated the so-called ``forced'' part of the solution
(with the scaling exponent $\gamma$) from the zero mode (that is constant in this 
case). It is the forced part that contributes the second order structural function.
The separation for ``forced'' terms and zero modes is valid for
higher order correlation functions as well.
It has been recognized independently by the authors of
\cite{95CFKLb,95GK,95SS} that there are zero modes ${\cal Z}$ that may
provide for an anomalous scaling. 
A zero mode, possessing the slowest downscale
decrease among the ones built on $2n$ points
(that is not reduced to a sum of zero modes each built on a less
number of points),
gives the major contribution into 
the $2n$-th order structural function of the scalar, for $n>1$ \cite{95CF}.
Scaling of such a zero mode should grow with $n$ to provide the convexity
of $\zeta_{2n}$ as a function of $n$ (it is an immediate
consequence of the Holder inequality, see for example \cite{95Fri}).
There are two Gaussian limits where there is no anomaly and it is easy to
make a classification of zero modes of operator $\hat{\cal L}$ there: 
a) limit of large space dimensionality, $d=\infty$;
b) so-called ``diffusive'' limit of the smooth scalar field, $\gamma=2^-$, 
(to be precise it was done even in a more restrictive case,
when $D/(2-\gamma)$ is finite). It was a recent breakthrough in the
analytic theory of turbulence, when the anomalous exponent $\zeta_{2n}$
was calculated perturbatively in the leading non-Gaussian order
in the respective small parameters:
$1/d$ in \cite{95CFKLb,95CF} and $2-\gamma$ in \cite{95GK,96BGK}.
One emphasizes that both the perturbative techniques do not work for sufficiently
large moments $2n$, when the anomalous corrections are of the order of
the normal scaling exponent $n\gamma$. To deal with  
$\zeta_{2n}$ for the largest moments, 
we shall deliver a nonperturbative instanton technique.

The basic equation (\ref{1a}) can be rewritten in the following evolution
form (one step back from the derivation of (\ref{1a}) presented in \cite{95CFKLb})
\begin{eqnarray}&&
F_{1\cdots 2n}=\int_0^\infty dT\exp\left[
T\left(-
\frac{1}{2}{\cal K}_{ij}^{\alpha\beta}\nabla_i^\alpha\nabla_j^\beta
+\kappa\sum_i\triangle_i\right)
\right]\chi_{1\cdots 2n}=\nonumber\\&&
\int_0^\infty dT \int \prod_i
d{\bf R}_i{\cal R}\{T;{\bf r}_i,{\bf R}_i\}
\chi_{1\cdots 2n}\{{\bf R}_i\},
\label{3}
\end{eqnarray}
here and everywhere below summation over the repeated particle and 
dimensional indexes will be assumed.
${\cal R}\{T;{\bf r}_i,{\bf R}_i\}$ is the resolvent of the
operator $\hat{\cal L}_{2n}$,
\begin{equation}
\left(\partial_t-\hat{\cal L}_{2n}
\{{\bf r}\}\right){\cal R}\{t;{\bf r}_i,{\bf R}_i\}=
\delta(t)\prod_i\delta({\bf r}_i-{\bf R}_i).
\label{R}
\end{equation}
Considering the differential operator under the exponent from the first line
of (\ref{3}) as
a Hamiltonian of a $2n$-particle Quantum mechanics we can rewrite the
resolvent in the Hamiltonian form of the standard Feynman-Kac path integral
\begin{eqnarray}&&
{\cal R}(T;{\bf r},{\bf R})=
\int_{{\bbox \rho}_i(0)={\bf r}_i}^{{\bbox \rho}_i(T)={\bf R}_i} 
\prod_i^{2n}{\cal D}{\bbox \rho}_i(t) {\cal D}{\bf p}_i(t)
\exp\left[-{\cal S}\{{\bbox \rho}(t);{\bf p}(t)\}\right],
\label{4a}\\&&{\cal S}=
\int_0^Tdt\left\{\frac{1}{2} 
 p_i^\alpha\left[{\cal K}\right]^{\alpha\beta}_{ij}p_j^\beta-
p_i^\alpha\dot{\rho}_i^\alpha\right\}
,\label{4b}
\end{eqnarray}
where $\hat{[\cal K]}$ is defined as:
\begin{equation}
[{\cal K}]^{\alpha\beta}_{ij}={\cal K}^{\alpha\beta}(\bbox{\rho}_i-
\bbox{\rho}_j)-2\kappa\delta^{\alpha\beta}\delta^{ij}.
\label{inv}
\end{equation}
Retarded regularization of the ``mass'' ($\hat{[\cal K]}$-term) in the 
action is considered, that means
the following discretization procedure:
$t_k=k\epsilon$, $\epsilon=T/M$, $k=0,\cdots,M$, 
 ${\cal D} {\bbox \rho}_i(t)=\prod_{k=1}^{M-1}d{\bbox \rho}_i(t_k)$,
${\bbox \rho}_i(t_0)={\bf r}_i$, ${\bbox \rho}_i(t_M)={\bf R}_i$, 
${\cal D} {\bf p}_i(t)=\prod_{k=1}^M d{\bf p}_i(t_k)$,
$M\to\infty$,
\begin{equation}
{\cal S}=\sum_{k=0}^{M-1}\left[\frac{\epsilon}{2}
p_i^\alpha(t_{k+1})[{\cal K}]_{ij}^{\alpha\beta}(t_k)p_j^\beta(t_{k+1})-
p_i^\alpha(t_{k+1})
\left(\rho_i^\alpha(t_{k+1})-\rho_i^\alpha(t_k)\right)\right],
\label{5}
\end{equation}
where the path integral for the associated Quantum mechanics could be understood
as explaining a random (Brownian) 
motion of $2n$ particles possessing a very special
dependence of the tensor of inverse masses $[\hat{\cal K}]$ on displacements of all
the particles. 
The resolvent represents the probability for the $2n$ fluid particles to diffuse from
the initial geometry ${\bf r}_i$ to the final one ${\bf R}_i$ for time $T$.
Notice that another $2n$-particle representation \cite{93Maj}
was used to analyze the pumping-free (decaying turbulence) two-dimensional
case of a linear ($\gamma=0$) anisotropic velocity profile.  

The representation (\ref{4a}-\ref{5}) is useless
if we aim to calculate the functional integral explicitly: 
it would reduce one
to calculation of the resolvent of $\hat{\cal L}$,
that is already stated as a generally unsolved problem.
Our aim is modest, we are going to study the higher correlation functions,
or many-particles problem ($n\gg d$) on the language of a ``quasi-classical''
approximation for the associated Quantum mechanics. 
The large parameter should allow us to evaluate the path integral from the
integrand of (\ref{4a}) (or its spatial derivatives, see below) in 
a saddle-point (instanton) manner.

$F_{2n}$ is not scale invariant.
The integrations over ${\bf R}$ and $T$ in (\ref{3}) give rise
to a huge set of zero modes, describing not only the $2n$-th structural function
but all the lowest ones too
(for details of the zero mode ideology see \cite{95CFKLb,95GK,95CF,96BGK}).
To separate zero mode giving the dominant contribution into 
the $2n$-th structural function, 
which is subleading in the zoo of the zero modes, 
we suggest an another oblique way of solving the problem.
The idea is to use an exact scaling relation between 
the $n$-th order moment of the dissipation field $\varepsilon=\kappa[\nabla\theta]^2$
and $2n$-th order structural function of the scalar
that was proved in \cite{95CF}
by means of the ultraviolet fusion rules discovered in \cite{95CFKLb}:
\begin{equation}
\mbox{if}\quad \langle(\theta_1-\theta_2)^{2n}\rangle
\sim r_{12}^{n\gamma}(L/r_{12})^{\Delta_{2n}}, \quad \mbox{then}\quad
\langle\varepsilon^n\rangle\sim (L/r_d)^{\Delta_{2n}},
\label{ep1}
\end{equation}
if it is known additionally that $\langle\varepsilon^n\rangle$ is scale-invariant.
Let us emphasize that the relation (\ref{ep1}) between structural function and
respective correlator of $\varepsilon$ is based crucially
on the expected scale-invariance of both the objects.
At $n$, considered to be the largest number in the theory,
the scale-invariance over $r/L$ does not need to be a priory valid.

We will construct an instanton for the correlator
of the dissipation field itself
\begin{eqnarray}&&
\langle\varepsilon^n\rangle\!\sim\!\kappa^n\lim_{r_i\to 0}\left[
\int_0^\infty\! dT\!\int \prod_i d{\bf R}_i
{\cal R}_\varepsilon\{T;{\bf r}_i,{\bf R}_i\}\chi_{2n}\{{\bf R}_i\}\right]
,\label{ep2a}\\&&
{\cal R}_\varepsilon\{T;{\bf r}_i,{\bf R}_i\}\!=\!
\int\limits_{{\bbox \rho}_i(0)={\bf r}_i}^{{\bbox \rho}_i(T)={\bf R}_i}
\prod_i^{2n}{\cal D}{\bbox \rho}_i(t){\cal D}{\bf p}_i(t) 
\prod_{k=1}^n\left[{\bf p}_{2k-2}(0)
 {\bf p}_{2k-1}(0)\right]
\exp\left[-{\cal S}\{{\bbox \rho}(t);{\bf p}(t)\}\right].
\label{ep2b}
\end{eqnarray}
It is easy to check by means of direct Gaussian
integrations that the discretization
condition (\ref{5}) reproduces the correct gradient structure of
the $\varepsilon$ correlator
(the Hamiltonian form of the path integral allows it to be easily checked).
A kind of pairing of the space indexes 
in the $p-p$ integrand of (\ref{ep2b}) is arbitrary
(for example, one could make the integrand symmetric with respect to all
permutations of all the particles).

There are two different specifications that we are free to fix
in the problem's set.
It concerns initial and final conditions imposed.
The initial condition 
is defined by the initial ${\bf r}_i$ geometry. 
The final condition is defined by the source function $\chi_{2n}$.
However, scaling exponents do not depend 
on a concrete form of the $\chi$-function
in accordance with the general zero-mode ideology \cite{95CFKLb,95GK,95CF,96BGK}.
One can use the freedom to make an appropriate choice of the 
initial geometry and the source function.

It is evident that integration over ${\bf R}_i$  in (\ref{ep2a})
cannot be performed
in the saddle-point manner, if the source function is, for example, 
a uniform constant
inside the circle $R<L$: All the values of ${\bf R}$ satisfied,
$R^\gamma\lesssim D T$
(the rough observation will be improved later on), 
give comparable weights in the integrand of (\ref{ep2a}). 
However, one can force a particular final geometry ${\bf R}_i\sim L$ 
to be preferable, choosing the source function to get a sharp maximum
about $R\sim L$, where $R$ is an average size, say 
$R=\sqrt{[\sum_i {\bf R}_i^2]/[2n]}$.
Then one can include the variation over ${\bf R}_i$ in the
common variation procedure adding  the term $-\ln[\chi_{2n}]$ to the action.
The formal trick is justified by
the general expectation to get the dominant contribution into
the $\varepsilon$ correlator from a zero mode of operator $\hat{\cal L}$.
It is the universal scaling of a zero mode that defines 
universal (independent on a concrete shape of $\chi_{2n}$)
scaling of the $\varepsilon$ correlator.

Thus, we are going to raise both the ${\bf p}-{\bf p}$ and source terms
from the integrand of (\ref{ep2b}) 
into the exponent to variate hereafter the effective action 
\begin{equation}
{\cal S}_{\mbox{eff}}={\cal S}-\sum_{k=1}^n\ln\left[
\kappa {\bf p}_{2k-2}{\bf p}_{2k-1}\right]-\ln[\chi_{2n}],
\label{eff}
\end{equation}
over all the allowed trajectories 
(over $\bbox{\rho}_i(t),{\bf p}_i(t)$ for all the $t$ from $0\leq t\leq T$)
in the next Section.

\section{Saddle-point approximation}

An instanton is defined from the saddle-point approximation - extremum
of the effective action (\ref{eff}) with respect to
fluctuating coordinates $\bbox{\rho}_i(t)$ and
momenta ${\bf p}_i(t)$ of all the $2n$ particles:
\begin{eqnarray}
&&\dot{p}_i^\alpha+
p_i^\beta p_j^\eta 
{\cal K}_{ij}^{\beta\eta;\alpha}=
\delta(T-t) \frac{\partial \ln[\chi_{2n}]}{\partial \rho_i^\alpha(T)}
,\label{6a}\\&&
\dot{\rho}_i^\alpha-[{\cal K}]^{\alpha\beta}_{ij}p_j^\beta=
-\frac{p_{i^*}^\alpha}{{\bf p}_j{\bf p}_{j^*}}\delta(t),
\label{6b}
\end{eqnarray}
where summations over the particle $j$ index and repeated spatial
indexes are supposed;
$j$ and $j^*$ are indexes of conjugated particles from a pair
(say $1$ and $2$ or $2n-1$ and $2n$); 
\begin{equation}
 {\cal K}^{\beta\eta;\alpha}(\bbox{\rho})\equiv\frac{\partial}{\partial\rho^\alpha} 
{\cal K}^{\beta\eta}(\bbox{\rho})=\frac{D}{\rho^\gamma}\left(
(d+1-\gamma)\rho^\alpha\delta^{\beta\eta}-\delta^{\alpha\beta}\rho^\eta-
\delta^{\alpha\eta}\rho^\beta+\gamma\frac{\rho^\alpha\rho^\beta\rho^\eta}{\rho^2}
\right).
\label{K}
\end{equation}
The discrete variant of the instanton equations is
\begin{eqnarray}
&&
\epsilon [{\cal K}]_{ij}^{\alpha\beta}(t_k)p_j^\beta(t_{k+1})+
\rho_i^\alpha(t_k)-\rho_i^\alpha(t_{k+1})=0,\label{dpk}\\&&
\epsilon p_i^\eta(t_{k+1}){\cal K}_{ij}^{\eta\beta;\alpha}(t_k)p_j^\beta(t_{k+1})+
p_i^\alpha(t_{k+1})-p_i^\alpha(t_k)=0,
\label{dr}\\&&
r_i^\alpha+\frac{p_{i^*}^\alpha(\epsilon)}{{\bf p}_i
(\epsilon){\bf p}_{i^*}(\epsilon)}=
\rho_i^\alpha(t_1),
\label{dp}
\\&& p_i^\alpha(t_M)=-\frac{\partial \ln[\chi_{2n}]}{\partial \rho_i^\alpha(t_M)},
\label{dR}
\end{eqnarray}
where $k$ temporal index in (\ref{dpk}-\ref{dr}) is running from $1$ to $M-1$;
$t_1=\epsilon=0^+$.
The equations (\ref{dp},\ref{dR}), appearing from
$\partial {\cal S}_{\mbox{eff}}/\partial p_i^\alpha(t_1)=0$,
and $\partial {\cal S}_{\mbox{eff}}/\partial \rho_i^\alpha(t_M)=0$ respectively, 
explain the rule of parameterization of the $\delta$-functions from the
rhs of (\ref{6b}) and (\ref{6a}). To study the saddle point trajectories at
the fixed initial geometry ${\bf r}_i$ and 
a fixed form of the source function $\chi_{2n}$
one should solve the following classical equations
of motion
\begin{eqnarray}&&
\dot{p}_i^\alpha+
p_i^\beta p_j^\eta 
{\cal K}_{ij}^{\beta\eta;\alpha}=0,\label{sem1}\\&&
\dot{\rho}_i^\alpha=[{\cal K}]^{\alpha\beta}_{ij}p_j^\beta,
\label{sem2}
\end{eqnarray}
in the boundary conditions 
\begin{equation}
{\bbox\rho}_i(0)={\bf r}^\prime_i,\quad
{\bbox\rho}_i(T)={\bf R}_i,\label{bc}
\end{equation}
where ${\bf r}^\prime_i\equiv{\bbox\rho}_i(\epsilon)$ is related to
${\bf r}_i$
and the initial momentum ${\bf p}_i(\epsilon\to 0)$ via (\ref{dp});
${\bf R}_i\equiv{\bbox\rho}_i(t_M)$ depends on the final momentum 
${\bf p}_i(t_M)$ via (\ref{dR}).
Therefore, the problem is reduced to resolving (\ref{sem1},\ref{sem2}) with
the boundary condition (\ref{bc}) fixed and
with the constraints (\ref{dp},\ref{dR}) imposed afterwards.

The classical Hamiltonian equations of motion (\ref{sem1},\ref{sem2})
possess the standard set of integrals of motion.
First of all, due to the independence of the Lagrangian
(the integrand part of the action ${\cal S}$) on time, the energy (that coincides
with the Lagrangian) is the conserved quantity
\begin{equation}
E=-\frac{1}{2}\dot{\rho}_i^\alpha p_i^\alpha
=\mbox{const}.
\label{E}
\end{equation}
Second, due to the invariance of the action with respect to a uniform
shift of all the particles, the momentum 
$P^\alpha=\sum_i p_i^\alpha$,
is conserved too. 
Choosing the appropriate coordinate frame, associated with
the center of inertia of the system of particles,
one forces the full momentum ${\bf P}$ to be equal to zero.
Third, due to the invariance of the action
with respect to uniform rotation (say, around the 
center of inertia, where $P^\alpha=0$) the angular momentum
\begin{equation}
M^{\alpha_3\cdots\alpha_d}=\rho_i^{\alpha_1}\epsilon^{\alpha_1\cdots\alpha_d}
p_i^{\alpha_2}=\mbox{const},\label{M}
\end{equation}
which is a $d-2$-dimensional antisymmetric tensor 
($\epsilon^{\alpha_1\cdots\alpha_d}$ is
the $d$-dimensional absolutely antisymmetric tensor), is
the last globally conserved quantity.

\subsection{Semi-classical analysis for correlators}

Let us consider a particular instanton solution describing
dynamical dispersion of particles from a geometry with 
$\rho_{ij}\sim r^\prime$ at the initial moment of time $0^+$,
to a final geometry of a common-type;
with at least one of the distances $\rho_{ij}(T)$ being of the order of $R\sim L$.
The major saddle-point contribution into the $\varepsilon$ correlator is
\begin{equation}
\langle\varepsilon^n\rangle^{cl}\sim\kappa^n\lim_{r\to 0} \left[
\int_0^\infty dT\exp[-S^{cl}_{eff}]\right],\label{es}
\end{equation}
where ${\cal S}^{cl}_{eff}$ 
is the effective action ${\cal S}_{eff}$ (\ref{eff}) taken on the
classical trajectory ${\bbox \rho}_i^{cl}$.
The law of temporal evolution of any such trajectory is
\begin{eqnarray}&&
\rho^{\gamma/2}-r^{\prime\gamma/2}\sim R^{\gamma/2} \frac{t}{T},\label{rp1}\\&&
p\sim \frac{R^{\gamma/2}}{nDT\rho^{1-\gamma/2}},\label{rp2}\\&&
{\cal S}^{cl}\sim \frac{nR^\gamma}{\alpha(n)TD},
\label{rp3}
\end{eqnarray}
where $0^+<t<T$.
The proportionality signs $\sim$ in (\ref{rp1}-\ref{rp3}) stand to point out that 
(\ref{rp1}-\ref{rp3}) are correct up to constant multipliers, depending on
the details of the geometry. Still, there are no extra scale and $n$ dependences 
in the explicit version of (\ref{rp1}-\ref{rp3}). 
The universal (with respect to the geometry's variation) scaling behavior,
$\rho\sim t^{2/\gamma}$, follows from equations
(any one of) (\ref{sem1},\ref{sem2}), by substituting there scaling 
(over time) ansatz for both $\rho$ and $p$ fields supported by the
energy conservation law (\ref{E}).
Here in (\ref{rp1}-\ref{rp3}),
considering all the distances to be much larger than $r_d$ we
dropped diffusion for a while. One accounts for diffusion 
in a special symmetrical case
(Appendix A1) aiming to show that to 
get the principal dependence of an ultraviolet
divergent quantity on $r_d$ it is enough, generally, 
just to replace all the separations
going to $0$ by $r_d$ (see also \cite{95CFKLb}).

We consider two very symmetrical cases in Appendix A:
1) the uniform expansion of a $S_m$ sphere 
($2n$ points uniformly distributed on the 
sphere), the $m\leq d$;
2) divergence of two drops, with $n_+$ and $2n-n_+$ particles merged in the
first and second points (drops) respectively.
One can separate all the possible trajectories on two different types
dependent on how $\alpha(n)$ behaves with $n$ going to $\infty$.
Most of trajectories, we will call them ``typical'', correspond to a
linear growth of $\alpha(n)$ with $n$. To specify,
the trajectory is ``typical'' if the volume bounded by a smooth $(d-1)$-dimensional
manifold built on the $2n$ points is not temporarily increased.
Relative divergence of the two-point geometry (see Appendix A3),
the same as expansion of the $S_m$ geometry with $m<d-1$, are ``typical''.
An example of ``non-typical'' trajectory is expansion of the $S_d$ sphere
(see Appendixes A1,A2 for an explanation of $S_m$ geometry).
$\alpha(n)/n$ decreases with $n$ going to $\infty$
for a ``non-typical'' trajectory.

For a specific kind of source function $\chi$ 
(possessing a sharp maximum) chosen, $R$ appears to be $\sim L$.
There are two different intervals over the integral time $T$. First,
$r^\prime=\rho(0^+)$ governed by (\ref{dp}) is about $r$ at
$0<T\ll L^{\gamma/2}r^{\gamma/2}/[D\alpha(n)]$; For the largest values
of the integral time,
$L^{\gamma/2}r^{\gamma/2}/[D\alpha(n)]\ll T$,
one gets $p(0)=1/r$ and $r^\prime\sim [rL^{\gamma/2}/(DT\alpha(n))]^{2/(2-\gamma)}$.
By substituting the saddle-point values, governed by (\ref{rp1}-\ref{rp3})
into (\ref{es}) one gets a divergence in the integral at the
largest times. The divergence is formal: it should be stabilized by 
the normalization factor, accounting for an algebraic decay of the resolvent with
$T$. We will see below (in Section IV) that the algebraic factor in fact
come into the game via account for fluctuations
to cut the temporal integration
in (\ref{es}) at $T\sim L^\gamma/[\alpha(n) D]$.
Thus, for all the values of $\gamma$ except some vicinities of the ``diffusive''
$\gamma=2$ and ``logarithmic'' $\gamma=0$ limits one gets
\begin{equation}
\langle\varepsilon^n\rangle^{cl}\sim [L/r_d]^{n\gamma}.
\label{ee}
\end{equation}
(\ref{ee}) accounts for the principal dependence of 
$\langle\varepsilon^n\rangle^{cl}$
on the Peclet number only.
It is the second interval (of the largest values) of $T$, that
give the dominant contribution in (\ref{ee}):
All the significant dependence on $r$ 
(or on $r_d$ after taking the limit in the rhs of (\ref{es})) in the
integrand of (\ref{es}) comes from the $p_0^{2n}$ term via
the multiplier $r^{2n}$. 
Finally, accounting for the dependence of $\kappa$ on the diffusion scale
results in the anomaly (\ref{ee}). 
Note, that the saddle-point result (\ref{ee}) 
is generic for all the space dimensions $d\geq 2$.

What is specific about some vicinity of $\gamma_d=2^-$ is an 
expected inapplicability
of the saddle point approximation there: A growth of ${\cal S}^{cl}_{eff}$ with
a growth in $n$ is diminished by decay of the action as $\gamma$ goes to $2^-$.
Note, that in the naive diffusion limit, $D=0$, (or in the special
limit $\gamma=2^-$, $D/(2-\gamma)=\mbox{const}$, see \cite{95GK}) 
a balance between different terms in (\ref{dp}) differs strongly 
from the general case: there the $r$-term can be dropped 
in comparison with $r$-independent ones.
$r^\prime$ turns out to be of the order of the pumping scale $L$,
that results in the absence of anomaly (as it should be:
the diffusion case is Gaussian).
An infinitesimally small deviation of $\gamma$ from $2^-$ in the special limit
of \cite{95GK} makes the $p_0\sim 1/r$ anomalous solution to be preferable
in comparison with the nonanomalous $p_0\sim 1/L$ one. Thus,
the $\gamma\to 2^-$ limit is indeed very peculiar.
The ``logarithmic'' $\gamma=0$ limit is very specific too.
What is written above in the general scheme is valid if 
$([r^\prime/L]^{\gamma/2}-[r^\prime/L]^\gamma)/\gamma\ll 1$ is
satisfied. However, the inequality ceases to be true at some very tight
vicinity of $\gamma=0$: The second term from the lhs of (\ref{dp})
could be dropped there
(the condition is opposite to the one which resulted in (\ref{ee})).

The anomalous answer (\ref{ee}) is generic: the scale invariance
holds true for the classical trajectories
of both ``typical'' and ``non-typical'' kinds. It accounts for 
the $n$-dependent prefactor in the integrand of 
(\ref{es}), discriminating between 
the two kinds of instantons.
At $R$ and $T$ considered to be fixed,
the $r$-independent term
of ${\cal S}^{cl}_{eff}$ gets no $n$-dependence 
in the first case of the ``typical'' instanton
(see Appendix A2,A3 for the two-point instanton and Appendix A1
for the $S_m$ spherical case with $m>d$).
Vice versa the ``non-typical'' instanton 
(it is a spherical case with $m=d$, for example) gets $n$-dependence
from the bare action (\ref{rp3}). $\alpha(n)$ goes to zero as $n$ goes to $\infty$
in this case. To conclude, 
the ``non-typical'' instanton ($S_d$ one) is suppressed in comparison
with the ``typical'' ones. However, we cannot distinguish between 
different ``typical'' instantons 
(the $S_m$ instanton with $m>d$ and the two-point instanton)
on the ``classical'' level.
The suppression of a ``non-typical'' instanton 
in $d$ dimensions has a clear physical
explanation. It follows from the conservation of the volume
of a fluid element, prescribed by the incompressibility condition.
It is a very rare trajectory (that means it has a small weight)
that stretches the $2n$ points forming the
$S_d$ sphere and conserves simultaneously the volume
enveloped by a $d-1$ surface built on the $2n$ points.
The surface cannot be smooth in this case, it is very fractal.

Note, that at $1<\gamma<2$ and $d=2$ there is not another symmetrical
instanton of the type discussed above except the $S_2$ one.
The two-point instanton that works pretty well at $0<\gamma<1$ turns out  
to be unstable at $\gamma>1$: particles being initially dropped together
into a group try to diverge from each other hereafter.
Most probably it is reasonable to study another symmetrical instanton
with all the particles being elongated into a straight line in this case.
Such an instanton could be preferable in comparison with the $S_2$ one.
We do not yet consider the straight-line instanton 
in the present paper, postponing it for a future study.

For any solution (already discussed or another) of the 
auxiliary problems (\ref{sem1},\ref{sem2},\ref{bc}) one can
design such an appropriate initial geometry ${\bf r}_i$ and
a particular form of the source function $\chi_{2n}$ that
the trajectory turns out to be a unique solution of the full
system of the saddle-point equations (\ref{6a},\ref{6b}):
Fixing ${\bf r}^\prime_i$ and ${\bf p}_i(0)$ one arrives at a unique
(due to constraint (\ref{dp})) initial geometry. Via explicit dynamics
and the second constraint (\ref{dR}), one finds an appropriate form
of the source function to make the saddle-point solution self-consistent.

The anomalous answer (\ref{ee}) is scale invariant. 
In the leading ``classical'' approximation
the anomaly is extraordinary large: The normal contribution into $\zeta_{2n}$
is fully compensated by the anomalous one, 
$\Delta^{cl}_{2n}\to n\gamma$ at $n\to\infty$.
However, the result gives no possibility for answering
the major questions : Is the $\varepsilon$ correlator 
scale-invariant at $n\to\infty$? And if it is scale-invariant,
what is the asymptotic of $\zeta_{2n}$ at the largest $n$?
To resolve the problems one should account for fluctuations.

It was established in the present Section
that the variety of saddle-point solutions
(for different ${\bf r}_i$ and $\chi$) gives the same scale dependence 
(\ref{ee}).
It supports the general statement \cite{95CFKLb,95CF}
that the dominant contribution into $\langle\varepsilon^n\rangle$ stems
from a scale-invariant zero mode of $\hat{\cal L}_{2n}$.
However, it follows from the same general statement too,
that the dominant contribution can be lacking
for a special kind of source function and initial geometry.
It is along this pathways that one should optimize the problem with respect to
${\bf r}_i$ and the $\chi$-function,
to find the dominant zero mode contribution:
We must not only find a contribution of fluctuations in
$\langle\varepsilon^n\rangle$ but show it is minimal.
	
\section{Account for Fluctuations}

Let us study in the path integral (\ref{ep2b}), Gaussian fluctuations 
about yet a nonspecified ``classical'' trajectory
(${\bbox \rho}^{cl}$ and ${\bf p}^{cl}$ are supposed to be known). 
The quadratic with respect to fluctuated fields
$\delta{\bbox \rho},\delta{\bf p}$, a correction to the classical
action ${\cal S}_{eff}^{cl}$ is
\begin{eqnarray}
&& \delta {\cal S}_{eff}\!=\!\frac{1}{2}\!\int_0^T\!dt
\left\{\!
\delta p_i^\alpha{\cal K}_{ij}^{\alpha\beta}\{{\bbox\rho}^{cl}\}
\delta p_j^\beta\!+\!
2\delta p_i^\alpha{\cal A}_{ij}^{\alpha\beta}\delta\rho_j^\beta\!+\!
\delta\rho_i^\alpha{\cal B}_{ij}^{\alpha\beta}\delta\rho_j^\beta\!-\!
2\delta p_i^\alpha\delta\dot{\rho}_i^\alpha\right\}\nonumber\\&&
-\frac{1}{2}\delta p_i^\alpha(0){\cal G}_{ij}^{\alpha\beta}\delta p_j^\beta(0),
+\frac{1}{2}\delta\rho_i^\alpha(T){\cal C}_{ij}^{\alpha\beta}
\delta\rho_j^\beta(T),
\label{dS}\\&&
{\cal A}_{ij}^{\alpha\beta}=
\delta_{ij}\sum_k{\cal K}_{ik}^{\alpha\nu;\beta}\{{\bbox\rho}^{cl}\}
{p^{cl}}_k^\nu-
{\cal K}_{ij}^{\alpha\nu;\beta}\{{\bbox\rho}^{cl}\}
{p^{cl}}_j^\nu,\nonumber\\&&
{\cal B}_{ij}^{\alpha\beta}=
\delta_{ij}\sum_k
{p^{cl}}_i^\nu{\cal K}_{ik}^{\nu\mu;\alpha\beta}\{{\bbox\rho}^{cl}\}
{p^{cl}}_k^\mu
-{p^{cl}}_i^\nu{\cal K}_{ij}^{\nu\mu;\alpha\beta}\{{\bbox\rho}^{cl}\}
{p^{cl}}_j^\mu
,\label{AB}
\\&&{\cal C}_{ij}^{\alpha\beta}=
-\frac{\partial^2 \ln[\chi_{2n}]}{\partial \rho_i^\alpha\partial \rho_j^\beta}
\biggr|_{\bbox{\rho}^{cl}(T)},\quad
{\cal G}_{ij}^{\alpha\beta}=-\frac{\delta^{\alpha\beta}\delta_{j i^*}}
{{\bf p}_i^{cl}(0){\bf p}_{i^*}^{cl}(0)},
\label{C}
\end{eqnarray}
where there is no summation over repeated $j$-indexes in (\ref{AB});
the pair of the particles' indexes $i$ and $i^*$ describe 
a conjugated pair of particles;
${\cal K}_{\bf r}^{\alpha\beta;\nu}$ and ${\cal K}_{\bf r}^{\alpha\beta;\nu\mu}$
stand for the first and second spatial derivatives of 
${\cal K}_{\bf r}^{\alpha\beta}$
(see explicit expression for the first derivative (\ref{K})).
Performing the Gaussian integrations over 
$\delta{\bbox \rho},\delta{\bf p}$ one arrives at the following
expression for the $\varepsilon$ correlator accounting for the Gaussian
fluctuations
\begin{eqnarray}&&
\langle\varepsilon^n\rangle\!\sim\!
\lim_{r\to r_d}\left[\int_0^\infty dT \exp\left[-{\cal S}_{eff}^{cl}\right]
{\cal Z}^{fl}\right],
\quad
{\cal Z}^{fl}=\biggl\langle\Psi_0\biggl|\Psi_T\biggr\rangle,
\label{ef}\\&&
\biggl\langle\Psi_0\biggl|=
\biggl\langle\prod_i\delta(\tilde{\bf r}_i)\biggl|
\exp\left[\frac{1}{2}{\cal G}_{ij}^{\alpha\beta}
\nabla^\alpha_{{\bf r}_i}\nabla^\beta_{{\bf r}_j}\right]
,\label{Psi0}\\&&
\biggl|\Psi_T\biggr\rangle=
\tilde{T}\mbox{exp}\left[\int_0^T 
\delta\hat{\cal L}dt\right]\biggr|
\delta\chi(\tilde{\bf r})
\biggr\rangle,
\quad \delta\chi(\tilde{\bf r})=\exp\left[
-\frac{1}{2}\tilde{\bf r}_i^\alpha{\cal C}_{ij}^{\alpha\beta}
\tilde{\bf r}_j^\beta\right],
\label{PsiT}\\&&
\delta\hat{\cal L}\{t;\tilde{\bf r}\}=-
\frac{1}{2}\sum_{i,j}\left\{{\cal K}_{ij}^{\alpha\beta}\{{\bbox\rho}^{cl}\}
\nabla^\alpha_{\tilde{\bf r}_i}\nabla^\beta_{\tilde{\bf r}_j}+
2{\cal A}_{ji}^{\beta\alpha}\tilde{r}_i^\alpha
\nabla^\beta_{\tilde{\bf r}_j}-
\tilde{r}_i^\alpha[{\cal B}]^{\alpha\beta}_{ij}\tilde{r}_j^\beta
\right\},\label{dL}\\&&
[\hat{\cal B}]=\hat{\cal B}-\hat{\cal A}^T
\hat{{\cal K}}^{-1}\hat{\cal A}.
\label{BC}
\end{eqnarray}
Here in (\ref{ef}), 
we use the canonical Quantum mechanical notations for matrix elements.
The operator in (\ref{Psi0}) is the descendant of the momentum's term
from the integrand of (\ref{ep2b}); the diffusive-like state (\ref{Psi0})
is well defined.
$\tilde{T}\mbox{exp}$ in (\ref{PsiT})
stands for an antichronological ordered exponential.
Thus, we came full circle at this stage of the calculations, returning back
to a problem in the operator representation form (compare the time-ordered
exponential from (\ref{PsiT}) with the original operator exponent, 
say from the first line of (\ref{3})). 
It follows from (\ref{PsiT}) that $\Psi_T$ can be understood as
a solution for the following differential equation
\begin{equation}
\left[\partial_t+\delta\hat{\cal L}\right]\Psi(t;\tilde{\bf r})=
\delta(t-T)\delta\chi(\tilde{\bf r}),
\label{deq}
\end{equation}
at an initial moment of time $\Psi_T(\tilde{\bf r})=\Psi(0;\tilde{\bf r})$.

Let us consider fluctuations about a ``typical'' saddle-point trajectory
with all the distances stretched somehow similarly
(we will specify the concrete form of the considered instantons later on). 
Performing rescaling of temporal and spatial variables in (\ref{deq})
one simplifies it. In the new dimensionless $\tau,{\bf s}_i$ variables
\begin{equation}
\tau=\frac{R^{\gamma/2}}{T}\int_0^t\frac{dt}{\rho^{\gamma/2}}=
\frac{\gamma}{2}\ln[\rho/r^\prime],\quad
{\bf s}_i=\frac{\sqrt{D}R^{\gamma/4}\tilde{\bf r}_i}{\sqrt{T}\rho^{1-\gamma/4}},
\label{taus}
\end{equation}
where $\rho(t)$ is a typically stretched (\ref{rp1}) classical trajectory,
(\ref{deq}) gets the following refined form
\begin{eqnarray}&&
\left[\partial_\tau+\hat{\cal L}^\prime\right]\Psi(\tau;{\bf s}_i)
=\delta\left(\tau-\frac{\gamma}{2}\ln[R/r^\prime]\right)
\delta\chi({\bf s}_iR^{1-\gamma/2}\sqrt{T/D}),\quad
0\leq\tau\leq\ln[R/r^\prime],
\label{L1}\\&&
\Psi_T=
\Psi\left(0;{\bf s}_i=
\sqrt{D/T}R^{\gamma/4}\tilde{\bf r}_i/r^{\prime 1-\gamma/4}\right),
\label{L1a}\\&&
\hat{\cal L}^\prime=
\frac{1}{2}\left\{-\tilde{\cal K}_{ij}^{\alpha\beta}
\nabla^\alpha_{{\bf s}_i}\nabla^\beta_{{\bf s}_j}+
2\tilde{\cal A}_{ji}^{\beta\alpha}s_i^\alpha
\nabla^\beta_{{\bf s}_j}-
s_i^\alpha\tilde{\cal B}^{\alpha\beta}_{ij}s_j^\beta
\right\}-\frac{4-\gamma}{2\gamma}s_i^\alpha\nabla_i^\alpha,
\label{L2}\\&&
\tilde{\cal K}_{ij}^{\alpha\beta}=\frac{1}{D\rho^{2-\gamma}}
{\cal K}_{ij}^{\alpha\beta}\{{\bf \rho}^{cl}\},
\quad \tilde{\cal A}_{ji}^{\beta\alpha}=
\frac{T\rho^{\gamma/2}}{R^{\gamma/2}}{\cal A}_{ji}^{\beta\alpha},\quad
\tilde{\cal B}^{\alpha\beta}_{ij}=\frac{DT^2\rho^2}{R^\gamma}
[{\cal B}]^{\alpha\beta}_{ij},
\label{L5}
\end{eqnarray}
where all the new dimensionless 
matrixes $\tilde{\cal K},\tilde{\cal A},\tilde{\cal B}$
are time ($\tau$) independent.
If we exclude divergent degrees of freedom
(we should worry about uniform rotation of the classical trajectory,
that is a soft mode, separately) from
$\hat{\cal L}^\prime$, it becomes a Hamiltonian of a well posed
Quantum Mechanics. It is the Quantum Mechanics
of $l$ Gaussian oscillators, $l\lesssim 2nd$. 
There is thus seen to be a gap in the spectrum of  
the reduced operator.
There are two essential (for present consideration) characteristics of
the energy spectrum: the value of the gap $\Delta_E$
and the level spacing $\delta_E$ between the ground state and the lowest
excited state.
Both the energetic characteristics are
positive functions of $n,d,\gamma$.
Stretching time $\tau=\gamma\ln[L/r^\prime]/2$ 
is a big parameter due to $Pe\gg d$.

There exist two different situations depending on how
$\delta_E$ behaves at the largest $n$.
First, $\tau\delta_E$ is a 
large parameter if $\delta_E$ does not decay as $n$ grows. 
Then, it is the evolution of the ground state 
$\Psi_{gr}\{{\bf s}_i\}$ giving the major contribution into $\Psi_T$,
\begin{equation}
\Psi_T=\Psi_{gr}
\{\sqrt{D/T}R^{\gamma/4}\tilde{\bf r}_i/r^{\prime 1-\gamma/4}\}
\left(\frac{r^\prime}{R}\right)^{\gamma\Delta_E/2}
\biggl\langle\Psi_{gr}\{{\bf s}_i\}
\biggl|\delta\chi({\bf s}_iR^{1-\gamma/2}\sqrt{T/D})\biggr\rangle 
.\label{PT}
\end{equation} 
The multiplier ${\cal Z}^{fl}$ 
is getting smaller with $r$ algebraically
\begin{eqnarray}&&
{\cal Z}^{fl}_{c-t}=
\int\prod_i d\tilde{\bf r}_i
\frac{\exp\left[-\tilde{r}_i^\alpha
 [\hat{\cal G}^{-1}]_{ij}^{\alpha\beta}\tilde{r}_j^\beta/2
\right]}{\sqrt{|\det[\hat{\cal G}]|}}\Psi^{(1)}_T\sim
\left[\frac{r^\prime}{L}\right]^{\gamma\Delta_E/2}
\left[\frac{r^{\prime 1-\gamma/4}\sqrt{T}}{rL^{\gamma/4}\sqrt{D}}\right]^l
\nonumber\\&&\sim
\left(\frac{r}{L}\right)^{[l/2+\Delta_E]\gamma/[2-\gamma]}
\left[\frac{n^{2-\gamma/2}T}{DL^\gamma}\right]^{-l/[2-\gamma]},
\label{Zflf}
\end{eqnarray}
where a typical matrix element of 
$[\hat{\cal G}^{-1}]_{ij}^{\alpha\beta}$ is estimated as
$[p^{cl}(t=0)]^2\sim r^{-2}$,
$T$ is considered to be smaller than $L^\gamma/D$, and 
$l$ counts the number of the stretched degrees of freedom. 
The second possibility is realized if $\delta_E$ is getting smaller 
with $n$ going to $\infty$. Hence it follows that 
one gets an evolution of a mixed wave packet
built from some amount of the lowest eigenstates: The wave function of
the ground state $\Psi_{gr}$ in (\ref{PT}) 
must be replaced by the wave function of the packet.
However the multiplier ${\cal Z}^{fl}$ is algebraic again and
the characteristic size of the packet has the same parametric dependence
as before. The parametric estimation (\ref{Zflf}) thus remains intact.
The number of the stretched degrees of freedom and the value of the gap
need to be specified in (\ref{Zflf}) to describe the anomaly
quantitatively.

We start the quantitative analysis
from discussion of the two-point instanton that is realized
at $1>\gamma>0$ only. It is an example of an instanton of the first type
with both $\delta_E$ and $\Delta_E$ being of the order of unity
(with respect to $n$). 
The dynamics of the two-point point instanton is characterized by
the relative divergence of the points (drops) along with a simultaneous
contraction of the sizes of the drops (see Section IIIA and Appendix A3). 
This means 
there are three different types of fluctuation degrees of freedom in this case.
First of all they are longitudinal fluctuations of the stretched degree of freedom;
second, fluctuations of the points (whole drops) 
in the $d-1$ directions transversal to
the stretched one; and, third, fluctuations of all the rest $(2n-1)d$
contracted degrees of freedom (intrinsic fluctuations of the drops).
One can calculate relative fluctuations of the drops and inner fluctuations of
the drops themselves independently
(it is easy to check afterwards that nondiagonal terms are negligible). 
Gaussian integrations account for relative 
longitudinal fluctuations of the points forms 
\begin{equation}
{\cal Z}^{fl}_{str} \sim[r/L]^{[1/2+\Delta_E^{st}]\gamma/[2-\gamma]}
\left[\frac{n^{2-\gamma/2}T}{DL^\gamma}\right]^{-1/[2-\gamma]},
\label{Z2}
\end{equation}
with $\Delta_E^{st}$ as calculated in Appendix B1.
One finds that the $d-1$ transversal fluctuations cannot be considered
as Gaussian ones (attempts to restrict their study by a Gaussian level leads to
divergence, see Appendix B1). 
Hopefully, one can calculate the transversal Non-gaussian
fluctuations explicitly. 
First, accounting for the relative fluctuations of the points (drops), 
at the initial ($t=0^+$) and final geometries fixed,
is performed by the method described in Appendix C.
Second, one can account for the transversal fluctuations of ${\bf R}_i$ and
${\bf r}_i^\prime$ explicitly too, 
calculating a variety of rotating classical trajectories with 
$T$, $R$ and $r^\prime$ taken from the rotationless trajectory,
whereas the angular momentum (\ref{M}) is nonzero
(the trajectories are found from
the auxiliary classical problem (\ref{sem1},\ref{sem2}) but not from the
full one (\ref{6a},\ref{6b}), with a fixed form of the source function
corresponding to the rotationless configuration). 
As a result, the trajectories with
nonzero angular momentums give the same value of the classical
action in the leading order in $(r^\prime/L)^\gamma$, 
as for the directly stretched rotationless case
(see explicit calculation for $d=2$ in Appendixes A).
In brief, accounting for the
contribution of strongly Non-Gaussian transversal fluctuations, results
in the $r$-independent multiplier (volume of the angular group).
We discuss fluctuations of the drops themselves 
(the rest $(2n-1)d$ fluctuating degrees of freedom) in 
Appendix B2. The fluctuations are short-correlated, that
results in $r$-independence of the respective contribution 
${\cal Z}^{fl}_\delta$ into ${\cal Z}^{fl}$.
However, the contribution (\ref{Zflfp}) 
shows an essential dependence on both $T$ and $n$.
Making substitution of (\ref{Z2}), (\ref{Zflfp}) 
(${\cal Z}^{fl}={\cal Z}^{fl}_{str}{\cal Z}^{fl}_\delta$),
and (\ref{rp3}) into (\ref{ef}),
and performing the integration over $T$ in the saddle-point manner,
one finally gets that characteristic value of the integral time
is getting smaller, $T\sim L^\gamma/[nD]\ll L^\gamma/D$, with $n\to\infty$.

It should be stressed once again 
that the result derived for $0<\gamma<1$ is not a
consequence of a specially chosen initial geometry and source term -
it is generic. 
For a majority of appropriate initial geometries and source functions, there is a 
special optimal configuration 
(that still may be difficult to find)
of values and orientations of the initial momenta,
making one distance diverge but all the rest converge
dynamically.

As it is shown in Section IIIA and Appendix A3, at $\gamma>1$ there exist 
no alternatives to the common-type stretching of all the degrees
of freedom: contraction of any distance (merging of any
particles in a point) leads to a singularity, that is forbidden.
And yet among all the generally stretched
instantons it is preferable to get ones characterized by 
a stretching with at least one
direction (dimension) kept stretching free (contracted):
This is what we call ``typical'' instanton.
``Typical'' symmetrical instantons  explained in Section IIIA and
Appendix A3  are $S_m$ spherical instantons
with $m<d$. Thus, let us apply the conducted above analysis for the case.
The $n$-dependence of the
potential term from the $2n$ particle Quantum mechanics (\ref{L2})
is estimated as $\tilde{B} _{ij}\sim 1/n^2$ for all the values of 
the particle index $j$, except
of ones from a small vicinity (on the sphere) of 
$i$ (every momentum is proportional 
to $1/n$ in the dimensionless variables).
For $i$ and $j$ being the 
nearest neighbors on the $S_m$ sphere
one gets $[B]_{ij}\sim n^{\gamma/[m-1]-2}$. One can drop all the
terms beside the nearest neighbors if 
$\gamma/[m-1]>1$. Vice versa,
if $\gamma/[m-1]<1$ one can replace all the matrix elements
by $\sim 1/n^2$ terms.
All the kinematic matrix elements ${\cal K}_{ij}$ are $n$-independent
(strictly speaking for $i$ and $j$ being close to each other
the matrix elements are even getting smaller with $n\to\infty$ than a constant, 
$\sim n^{-[2-\gamma]/[m-1]}$). Hence at $\gamma/[m-1]<1$
the energy characteristics are estimated
as $\delta_E\sim n^{-1}$, and $\Delta_E$ as a constant respectively.
In the opposite case $\gamma/[m-1]>1$ (that
is realized only if $m=2,\gamma>1$) one gets
$\delta_E\sim n^{\gamma/[2(m-1)]-1}$, $\Delta_E\sim n^{\gamma/[2(m-1)]}$.
Calculation of the $n$-dependence of the ${\cal A}$ term does not
change the principal dependence on $n$ of the energy characteristics.
There is the extra parameter $l$ which enters the anomalous answer
and counts the number of typically stretched degrees of freedom.
For the $S_m$ instanton one gets $l=2(m-1)n$.
To conclude, contribution of fluctuations about 
the $S_m$ instantons, $d-1>m\geq 2$, is estimated as
$\sim [r/L]^{(\Delta_E+(m-1)n)\gamma/[2-\gamma]}
[n^{2-\gamma/2}T/[DL^\gamma]]^{-2(m-1)n/[2-\gamma]}$. 
By this means the Gaussian correction appears to be of the same order as
(even larger than)  the saddle-point value (\ref{ee}),
rendering the $S_m$ saddle-points to be smooth out by the Gaussian fluctuations.
Particularly, at $\gamma>1$ the contribution of the 
$S_m$ instanton (saddle-point $+$
Gaussian fluctuations) into the 
$\langle\epsilon^n\rangle$ correlator is getting smaller with $\mbox{Pe}$
increase. The contribution is of no interest since it is
negligible in comparison with the forced term contribution
(possessing the normal scaling) that was dropped in the saddle-point approach
from the very beginning.
One recognize that 
the saddle-point calculus is not an appropriate tool
for calculations of the anomalous exponent at $\gamma>1$.

It was thus shown in the present Section
that at $0<\gamma<1$ the contribution of Gaussian fluctuations into
$\langle\varepsilon^n\rangle$ is algebraic (scale-invariant)
with respect to the Peclet number ($\mbox{Pe}=L/r_d$) and it is
small in comparison with the ``classical'' value (\ref{ee}).
The scaling exponent $\zeta_{2n}$ of 
the scalar's structural function shows a finite limit
at $n\to\infty$. The exponent is calculated explicitly
\begin{equation}
\mbox{at}\quad 0<\gamma<1,\quad
\zeta_\infty=\frac{\gamma}{2(2-\gamma)}+
\frac{d+2-\gamma}{2(2-\gamma)}\left[2-3\gamma/2+
\sqrt{16-16\gamma+17\gamma^2/4}\right].
\label{z2n}
\end{equation}
There are relative fluctuations of the points 
in the two-point geometry that are responsible for the answer (\ref{z2n}).
At $\gamma>1$ all the saddle-point solutions discussed in Section III
are smoothed out by the Gaussian fluctuations: The instanton calculus does not
work in this case.

\section{Conclusion}

It was stated in the Introduction that the idea of the saddle-point 
calculus is to make $n$ the largest number in the problem.
However, to establish the criterion of applicability of the saddle-point
approximation at $0<\gamma<1$ explicitly one should estimate 
contributions of Non-Gaussian fluctuations about
the instanton and to compare them with the already found Gaussian corrections.
If following the general scheme (\ref{ef}-\ref{L5}) to keep a nonlinear
(say, third order over $\tilde{r}$ term) one arrives at an extra factor
$\sqrt{T}/[R\rho(t)]^{\gamma/4}\ll \mbox{Pe}^{\gamma/4}/\sqrt{n}$
behind the dimensionless $\sim \tilde{s}^3$
term in the nonlinear variant of (\ref{L2}). 
The factor (along with the integral time $T$)
is getting smaller with $n\to\infty$
(the smallness makes the saddle-point become instant, 
and the ``classical'' action become, respectively, large).
The observation is generic: all the higher order corrections
to the Quantum mechanics describing the Gaussian fluctuations are small 
if $n$ is the largest number in the problem
\begin{equation}
n\gg \mbox{Pe}^{\gamma/2},d.
\label{crit}
\end{equation}

It is remarkable that the anomalous scale-invariant result (\ref{z2n}) 
has a more wider criterion of applicability than the method used for 
its derivation.
Indeed, it is the zero mode of the primary operator $\hat{\cal L}$ that
makes the major contribution into the $2n$-th order structural function.
Choosing the scale-invariance to be the key for classification
of zero modes, one finds that only $n$, $d$ and $\gamma$ (but not $\mbox{Pe}$)
can be entered into the zero modes. 
Thus, the anomaly (\ref{z2n}) is valid even if $\mbox{Pe}^{\gamma/2}$ 
is larger than $n$, however $n$ remains much larger than $d$.
%Strictly speaking, the anomaly (\ref{z2n}) forms
%the upper bound for $\zeta_{2n}$ at $\mbox{Pe}^{\gamma/2}\gg n\gg d$:
%a zero mode with a lower scaling exponent
%could be subleading in the calculations performed 
%at $n\gg\mbox{Pe}^{\gamma/2}\gg d$. Nonetheless, it is very unlikely physically.
It is important to note that (\ref{z2n}) matches 
parametrically correct the perturbative results
of the $1/d$ \cite{95CFKLb,95CF} expansion at $n$ being of the order of $d$.
 
There existed two different expectations for $\zeta_{2n}$ at 
$n$ being sufficiently large.
Decoupling of the molecular-diffusion term in the
equation (that is not closed originally) for the structural function of
the $2n$-th order gives $\zeta_{2n}\to\sqrt{2nd\zeta_2}$ \cite{94Kra,95KYC}.
Another prediction is that $\zeta_{2n}$ tends to a $n$-independent constant
determined somehow by $d$ and $\zeta_2$ \cite{94Fal,94Kra,95Yak}. 
Thus, the recent work \cite{95Yak} is based
on an extension onto the passive scalar case
of the method of operator product expansion
suggested recently in the context of Burgers' turbulence \cite{95Pol}.
Our calculations contradict the first prediction and support
the prediction for $\zeta_{2n}$ to approach a 
constant at $n\to\infty$ if $0<\zeta_2<1$.
It is worth noting that an extended prediction of 
\cite{95Yak} gives us even more than an asymptotic constant behavior
for the exponent at large $n$: $\zeta_{2n}$ was predicted to be a constant
for all the numbers $n$ having been larger than some $n_0$.
We cannot exclude or confirm the extended prediction here.
It will require accounting for corrections to the saddle-point solution 
of the next non-Gaussian order explicitly. 
Let us stress that the equivalent problem in the Burgers' 
turbulence is not yet solved too.

A comparison of the first-quantized formalism (``Quantum'' particles)
presented in the paper with a  
second-quantized one (``Quantum'' fields), that hopefully will be
developed in accordance with the general scheme \cite{95FKLM},
would be very instructive.

It was argued phenomenologically
\cite{95SS,96SS} and confirmed quantitatively at $d\gg n$ by means of
direct expansion over finite velocity's
correlation time \cite{96CFL},
that the scalar's exponents $\zeta_{2n}$ are nonuniversal,
they are crucially aware of the velocity's temporal characteristics.
In the present paper we developed a theory for a peculiarly adopted
(to the analytical study) case of the $\delta$-correlated 
Gaussian velocity field. Nevertheless, the starting technical idea of the
paper to replace $\nabla_i$ in the operator representation 
by momentum of $i$-th particle in the path integral does not
require any temporal or statistical restrictions on the velocity field
being imposed. Hence it follows that
it would be interesting to generalize the theory of finding
$\zeta_{2n}$ at $n\gg d$ for
the more realistic case of finite temporal correlations and generally for
arbitrary degrees of non-Gaussianity of the velocity field.
It remains to be seen whether
the scaling of the largest moments is nonuniversal in the limit.
It is hoped that quantitative comparison of the future theory
with experimental data 
(say, with measurements of temperature structural functions,
up to order $12$ \cite{84AHGA}; for review of experiments see
\cite{91Sre}) will be real some day.

The theory prediction does not contradict simulations
for $d=2$, $\zeta_2=1/2$ reported in \cite{95KYC}
(there, the velocity field was swept rapidly through the scalar to
mimic the short-correlated feature of the Kraichnan's model). 
The largest tenth moment measured in the simulation
gives $\zeta_{10}\approx 1.6085$, that is smaller 
(as it should be due to the convexity inequality) than our asymptotic
result $\zeta_\infty\approx 5.1$.

\acknowledgements
This work is a part of the extensive program on studying anomalous
scaling in turbulence undertaken together with 
E. Balkovsky, G. Falkovich, I. Kolokolov, and V. Lebedev
in the Weizmann Institute.
We strongly benefited from very helpful numerous discussions with them.
Useful remarks of A. Kamenev, D. Khmelnitskii, S. Levit, D. Orgad, 
and V. Yakhot are gratefully acknowledged.
This work was supported by the Clore Foundation.

\appendix
\section{Source-free symmetrical instantons}

\subsection{$S_l$ sphere geometry in $d$ dimensions}

Consider $2n$ points equidistantly distributed on the $l$ dimensional sphere:
for example in the two-dimensional case, $d=l=2$, we can use the following 
polar representation $\bbox{\rho}_k=(\rho,\varphi_k)$,
$\varphi_k=(k-1)\pi/n$, $k=1,\cdots,2n$, for the points' displacements. 
Let us consider the following setup 
1) the angular momentum (\ref{M}) to be equal to zero;
2) the initial ($t=0^+$) spherical geometry to be preserved dynamically.
Then, all the vector objects
defined for some particle $i$, like $\dot{\bbox{\rho}}_i$ or 
${\bf p}_i$, are parallel to $\bbox{\rho}_i$. The system of equations
(\ref{sem1},\ref{sem2}) is reduced to
\begin{eqnarray}
&& \dot{p}-a_2 Dp^2 \rho^{1-\gamma}=0,
\label{sc2a}\\
&& \dot{\rho}+a_1 Dp \rho^{2-\gamma}+2\kappa p =0,
\label{sc2b}\\
&& a_2=-\frac{1}{D\rho^{1-\gamma}} n_i^\alpha n_i^\beta \sum_j n_j^\eta 
{\cal K}^{\beta\eta;\alpha}(\bbox{\rho}_i-\bbox{\rho}_j)\nonumber\\&& =\frac{1}{2}
\sum_k\left(2\sin[\varphi_k/2]\right)^{2-\gamma}\left(
(2d-\gamma)\sin^2[\varphi_k/2]-d-1+\gamma\right)>0,
\label{a1}\\&&
a_1=-\frac{1}{D\rho^{2-\gamma}} n_i^\alpha \sum_j n_j^\beta
{\cal K}^{\alpha\beta}(\bbox{\rho}_i-\bbox{\rho}_j)=
\frac{2a_2}{2-\gamma}.
\label{a2}
\end{eqnarray}
Thus, we arrive at the simple equation with the following boundary
conditions imposed 
\begin{eqnarray}&&
\frac{d}{dt}\left[\frac{\dot{\rho}}{a_1D\rho^{2-\gamma}+2\kappa}\right]+
\frac{a_2 D\rho^{1-\gamma}\dot{\rho}^2}{(a_1D\rho^{2-\gamma}+2\kappa)^2}=0, 
\label{7a}\\&&
\rho(0)=r^\prime,\quad\quad \rho(T)=R.\label{7d}
\end{eqnarray}
Solution of the equation (\ref{7a}) fixed by the conditions (\ref{7d}) is
\begin{equation}
\int_{r^\prime}^{\rho(t)}\frac{dx}{\sqrt{a_1Dx^{2-\gamma}+2\kappa}}=\frac{t}{T}
\int_{r^\prime}^{R}\frac{dx}{\sqrt{a_1Dx^{2-\gamma}+2\kappa}},
\label{8a}
\end{equation}
in accordance with the energy conservation law (\ref{E})
(to get the answer (\ref{8a}) 
one could replace one of the basic equations (\ref{sc2a}-\ref{sc2b})
by the conservation law (\ref{E})). On the present instanton solution (\ref{8a})
the action (\ref{4b})
gets the following form
\begin{equation}
{\cal S}^{cl}(T;r^\prime,R)=\frac{n}{2T}\left(\int_{r^\prime}^R\frac{dx}
{\sqrt{a_1Dx^{2-\gamma}+2\kappa}}\right)^2.
\label{act}
\end{equation}
The momentum of the $k=0$'th particle at zero moment of time is
\begin{equation} 
p\biggl|_{t=0}=-\frac{ \int_{r^\prime}^R dx
[a_1Dx^{2-\gamma}+2\kappa]^{-1/2} }{T\sqrt{a_1Dr^{\prime2-\gamma}+2\kappa}}.
\label{pp}
\end{equation}

Calculating $a_1$ in the two-dimensional case of $S_2$ geometry,
one finds that
$a_1=[(4-\gamma)b_2/4-(3-\gamma)b_1]/(2-\gamma)$ and
\begin{eqnarray}&&
b_1=\sum_k(2\sin[\varphi_k/2])^{2-\gamma},\quad b_1\biggl|_{n\to\infty}\to
\frac{2^{4-\gamma}\Gamma[(3-\gamma)/2]}{\Gamma[(4-\gamma)/2]}n,
\label{b1}\\&&
b_2=\sum_k(2\sin[\varphi_k/2])^{4-\gamma},\quad b_2\biggl|_{n\to\infty}\to
\frac{2^{6-\gamma}\Gamma[(5-\gamma)/2]}{\Gamma[(6-\gamma)/2]}n.
\label{b2}
\end{eqnarray}
As $n$ goes to $\infty$ (and $\gamma$ being not too closed to $2^-$)
$a_1$ goes to zero as $n^{\gamma-2}$: If to replace $b_{1,2}$ in the definition
of $a_1$ by their asymptotic values (\ref{b1},\ref{b2}), a remarkable
cancelation, that gets rid of the linear over $n$ term in $a_1$, takes place.
The cancelation occurs as a direct consequence of the incompressibility
of the flow (the divergeless of the ${\cal K}$ matrix).
Thus, the major contribution into $a_1$ is estimated by the first 
term of the series over $n$,
\begin{equation}
a_1\biggl|_{n\to\infty}\to n^{-[2-\gamma]/[d-1]},
\label{aaa}
\end{equation}
where it is calculated that the angular size of the elementary cell,
appeared after sectioning of the $S_d$ sphere on $n$ parts, is
proportional to $\pi/n^{1/[d-1]}$.
Considering $S_2$ geometry in higher dimensions $d>2$,
and generally $S_l$ geometry in $d>l$, one finds
a linear growth of $a_1$ as $n$ goes to $\infty$.

\subsection{$S_2$ geometry. General two-dimensional case.}

Let us fix the initial and final vectors, that define the symmetric
geometry (say, $\bbox{\rho}_1(0)$ and $\bbox{\rho}_1(T)$),
to be nonparallel to each other. The situation starts to be 
more complicated, if one is going to solve the same
instanton equation (\ref{6a}), as new parameters
describing  rotation of the vector come into the game.
We consider here the two-dimensional case, that is the simplest one
(rotation is Abelian in this case). In $d=2$
the vectors $\bbox{\rho}_i$ and ${\bf p}_i$ can be parameterized as
$\rho_i^\alpha\to \rho\exp[i(\varphi+\varphi_k)]$,
$p_i^\alpha\to p\exp[i(\psi+\varphi+\varphi_k)]$,
where scalars $p(t)$, $\rho(t)$ and angles
$\varphi(t)$, $\psi(t)$ are time-dependent.
Four equations describing the dynamical behavior are
\begin{eqnarray}&&
\dot{p}\!+\!Dp^2\rho^{1-\gamma}
\left[(3-\gamma\cos^2[\psi])b_1+\frac{\gamma-4}{4}b_2\right]\frac{\cos[\psi]}{2}
\!=\!0,
\label{rn1}\\&&
\dot{\psi}\!+\!\dot{\varphi}\!+\!Dp\rho^{1-\gamma}
\left[(\gamma\cos^2[\psi]-1)b_1+\frac{4-\gamma}{4}b_2\right]\frac{\sin[\psi]}{2}
\!=\!0,
\label{rn2}\\
&&\dot{\rho}\!=\!\frac{Dp\rho^{2-\gamma}}{2-\gamma}\!
\left[(3-\gamma)b_1+\frac{\gamma-4}{4}b_2\right]\cos[\psi]-2\kappa p\cos[\psi],
\label{rn3}\\&&
\dot{\varphi}\!=\!\frac{Dp\rho^{1-\gamma}}{2-\gamma}
\left[b_1+\frac{\gamma-4}{4}b_2\right]\sin[\psi]-2\kappa\frac{p}{\rho}\sin[\psi],
\label{rn4}
\end{eqnarray} 
where $b_1$ and $b_2$ were introduced in (\ref{b1},\ref{b2}).
The equations of motion (\ref{rn1}-\ref{rn4}) are
compatible with the conservation laws of energy 
(\ref{E}) and the angular momentum (\ref{M}) (that is pseudoscalar in this case).
The system of equations (\ref{rn1}-\ref{rn4}) can be analyzed in full glory.
Nevertheless starting from the point one drops diffusion term responsible
for the ultraviolet regularization only.
Then in the diffusion-free case, the system of equations 
(\ref{rn1}-\ref{rn4}) being rewritten
in terms of the auxiliary $x,y$ variables, $(x,y)=\rho^{\gamma/2}
(\cos[\gamma\varphi/(2\sqrt{\eta})],\sin[\gamma\varphi/(2\sqrt{\eta})])$,
describes a uniform motion of a particle with constant speed in the $x-y$ plane:
\begin{eqnarray}&&
x(t)=r^{\prime\gamma/2}+\frac{t}{T}\left(
\cos[\gamma\varphi_*/(2\sqrt{\eta})]
\sqrt{R^\gamma-
2R^{\gamma/2}r^{\prime\gamma/2}\cos[\gamma\varphi_*/(2\sqrt{\eta})]+
r^{\prime\gamma}}-
r^{\prime\gamma/2}\right),\label{x}\\&&
y(t)=\frac{t}{T}
\sin[\gamma\varphi_*/(2\sqrt{\eta})]
\sqrt{R^\gamma-
2R^{\gamma/2}r^{\prime\gamma/2}\cos[\gamma\varphi_*/(2\sqrt{\eta})]
+r^{\prime\gamma}},\label{y}
\end{eqnarray}
where
\begin{equation}
\eta=\frac{[(4-\gamma)b_2/4-b_1]}
{[(4-\gamma)b_2/4-(3-\gamma)b_1]},
\label{eta}
\end{equation}
and the following initial and final conditions 
for $\rho,\varphi$ dynamical fields are
imposed: $\rho(0)=r^\prime$, $\rho(T)=R$ and 
$\varphi(0)=0$, $\varphi(T)=\varphi_*$.
The solution (\ref{x},\ref{y})
means, particularly, that there are no equivalent trajectories
for $\varphi_*$ taken from the interval 
$|\varphi_*|\leq 2   \pi\sqrt{\eta}/\gamma$. 
Thus, considering the final values of $\varphi_*$ and 
$\varphi_*+2\pi n$ from the interval
to be equivalent, one observes $\sqrt{\eta}/\gamma$-fold degeneracy
(all those trajectories from the degenerate set are differed by the values
of energy $E$ and angular momentum $M$). 

The action ${\cal S}$ on the ``classical'' trajectory gets the following form
\begin{equation}
{\cal S}^{cl}\!=\!
\frac{4(2-\gamma)n}{\gamma^2[(4-\gamma)b_2/4-(3-\gamma)b_1]D }
\frac{R^\gamma-2r^{\prime\gamma/2}R^{\gamma/2}\cos[\gamma\varphi_*/(2\sqrt{\eta})]+
r^{\prime\gamma}}{T}.
\label{S}
\end{equation}
The momentum of the $k=0$-th particle at zero moment of time
(the object entered the self-consistency condition (\ref{dp}) and
the pre-exponent term) has the following dependence on the initial and 
final conditions imposed: ${\bf p}_i(t=0)=p_0\exp[i\varphi(0)+\varphi_k+\psi_0]$,
\begin{eqnarray}&&
p_0\cos[\psi_0]\!=\!-\!\frac{2(2\!-\!\gamma)}{\gamma
[(4\!-\!\gamma)b_2/4\!-\!(3\!-\!\gamma)b_1]DT}\!
\nonumber\\&& \times
\left(\!\sqrt{
R^\gamma\!-\!2r^{\prime\gamma/2}R^{\gamma/2}\cos[\gamma\varphi_*/(2\sqrt{\eta})]\!+\!
r^{\prime\gamma}}\cos[\gamma\varphi_*/(2\sqrt{\eta})]\!-\!r^{\prime\gamma/2}\!
\right)\!
r^{\prime\gamma/2-1},\label{p1}\\
&& p_0\sin[\psi_0]=-\frac{2(2\!-\!\gamma)}{\gamma\sqrt{\eta}
[(4\!-\!\gamma)b_2/4\!-\!(3\!-\!\gamma)b_1]DT}\nonumber\\ &&\times\sqrt{
R^\gamma\!-\!2r^{\prime\gamma/2}R^{\gamma/2}\cos[\gamma\varphi_*/(2\sqrt{\eta})]\!+\!
r^{\prime\gamma}}\sin[\gamma\varphi_*/(2\sqrt{\eta})]r^{\prime\gamma/2-1}.
\label{p2}
\end{eqnarray}

 \subsection{Two-point geometry. $0<\gamma<1$}
The saddle-point system of equations (\ref{sem1},\ref{sem2}) 
has some reduction feature at $0<\gamma<1$:
if we merge a group of particles in a point
and choose the momenta (equal for the particles pasted together)
to be parallel to the vector connecting the points at the
initial moment of time, the problem gets rid of those
superfluous degrees of freedom at all the 
latest times too  - the group can be replaced by one particle.
We will check the general observation for the two-point
case in the present Appendix.

Here we consider a geometry formed by two groups 
(labeled by $+$ and $-$) of particles
($n_++n_-=2n$) each merged in a point with a position
$\bbox{\rho}_+(t)$ or $\bbox{\rho}_-(t)$ respectively. 
Indeed, there exists a solution of the saddle point equations (\ref{sem1},
\ref{sem2}) that preserves the symmetry dynamically. That is specific about
$\gamma<1$, it is an algebraic decay of ${\cal K}_r^{\alpha\beta;\eta}$
when $r$ goes downscales; one can put the particles in a point without
any divergences appearing. 
The number of dynamical degrees of freedom
is reduced from $4nd$ to $2d$. One gets 
\begin{eqnarray}&&
\dot{p}^\alpha_\pm+n_\mp p^\beta_+p^\eta_-
{\cal K}_\rho^{\beta\eta;\alpha}=0,
\quad \bbox{\rho}=\bbox{\rho}_+-\bbox{\rho}_-, 
\label{2p1}\\&&
\dot{\rho}^\alpha_\pm+2\kappa n_\pm p^\alpha_\pm
-n_\mp{\cal K}_\rho^{\alpha\beta}
p^\beta_\mp=0.
\label{2p2}
\end{eqnarray}
Coming from ${\bbox\rho}_\pm,{\bf p}_\pm$  variables to
collective ones
${\bf p}=n_+{\bf p}_+=-n_-{\bf p}_-$, ${\bbox\rho}$ 
(the full momentum is constant in the reference frame 
associated with the center of inertia), one gets the system of equations
that does not depend on the numbers of $+$ and $-$ particles at all
\begin{eqnarray}&&
\dot{p}^\alpha-p^\beta{\cal K}_\rho^{\beta\eta;\alpha}p^\eta=0, 
\label{2p3}\\&&
\dot{\rho}^\alpha+4\kappa p^\alpha
+2{\cal K}_\rho^{\alpha\beta}
p^\beta=0.
\label{2p4}
\end{eqnarray}
Notice that the rotationless variant of (\ref{2p3},\ref{2p4}) is transformed
to (\ref{sc2a},\ref{sc2b}) 
and the two-dimensional variant of (\ref{2p3},\ref{2p4}) is transformed
to (\ref{rn1}-\ref{rn4}),
if one performs a reduction
$a_1\to 2(d-1)/(2-\gamma)$, and $\kappa\to 2\kappa$ and $\eta\to \gamma^2/4$ there. 
Thus, the formulas for the ``classical'' action (\ref{act}) and
the initial momentum (\ref{pp})
are valid with appropriate
changes (the multiplier $[2n]^{-1}$ 
should be accounted additionally in the action)
in the case of the two-point geometry too.
One finds that the effective ``classical'' 
action (\ref{eff}) is $n$-independent in this case,
if the values of $T$ and $R$ are considered to be fixed.

To check the dynamical stability of the two-point geometry
let us consider a ``dumb-bell'' geometry
(floated variant of the two-point one) with the characteristic size of
the drops of the ``dumb-bell'' being initially (at $t=0^+$)
much smaller than the relative distance between them.
The major question to ask is: Will the ``dumb-bell''geometry be preserved
dynamically? To answer the question, let us go back
to the classical equation (\ref{sem2}). Introduce a small excursion
of a particle ${\bf \rho}_\delta$ from the drop's center.
Then, keeping the leading (over $\delta\rho/\rho$) term in the rhs of (\ref{sem2})
one arrives at the following equation for $\delta{\bbox\rho}$
\begin{equation}
\delta\dot{\rho}^\alpha=p \rho^{1-\gamma} \left(\delta\rho^\alpha-
d(\delta{\bbox\rho}{\bf n})n^\alpha\right).
\label{pp1}
\end{equation}
Substituting the already known law of the basic
two-point stretching, one gets,
\begin{eqnarray}&&
\frac{\delta\rho_\bot(t)}{\delta\rho_\bot(0^+)}=
\left(\frac{r^\prime}{\rho}\right)^{(2-\gamma)/[2(d-1)]},
\label{pp2}\\&&
\frac{\delta\rho_\|(t)}{\delta\rho_\|(0^+)}=
\left(\frac{\rho}{r^\prime}\right)^{1-\gamma/2},\label{pp3}
\end{eqnarray}
where $\delta\rho_{\bot,\|}$ are transversal and longitudinal 
(with respect to the direction of the ``classical'' stretching) sizes of the drops.
To conclude, the ratios of the drops' sizes to
the distances between the drops are getting smaller with time
and the two-point geometry is indeed preserved dynamically.
Note that (\ref{pp2}) is a ``classical''manifestation 
of the law of volume conservation
valid at $\gamma=0$ (the respective law of area conservation
allows solving the problem at $d=2$ explicitly \cite{95BCKL,95SS}): 
$\rho[\delta\rho_\bot]^{d-1}$ does not depend on time.
In other words, (\ref{pp2}) is a statistical descendant of the volume conservation
law (incompressibility condition) valid for any particular
flow.

Considering the dumb-bell geometry at $\gamma>1$, one
finds that it is destroyed dynamically:
The size of the drops and separation between them,
in the initially served two-point geometry, turns out
to be of the same order at the latest times.

\section{Gaussian fluctuations about
the two-point instanton}
\subsection{Relative fluctuations of the points}

Consider Quantum mechanics (\ref{L1}-\ref{L5}),
that appears when accounting for the relative fluctuations of the drops only.
There are $d$ essential distances in the case: 
$\tilde{\bf r}=\tilde{\bf r}_{\{1\}}-\tilde{\bf r}_{\{2\}}$.
Operator $\hat{\cal L}^\prime$ (\ref{L2}), 
rewritten in terms of ${\bf s}$ (related to $\tilde{\bf r}$
via (\ref{taus})), is 
\begin{eqnarray}&&
\hat{\cal L}^\prime\!=\!
\left[\frac{d+1-\gamma}{2-\gamma}\delta^{\alpha\beta}
\!-\!n^\alpha n^\beta\right]\nabla_{\bf s}^\alpha\nabla_{\bf s}^\beta\!+\!
\frac{1}{\gamma}\left[\!-\!\left(\frac{4-\gamma}{2}\!+\!
\frac{2(2-\gamma)}{d-1}\right)
s^\alpha\!+\!\frac{2(2-\gamma)d}{d-1} 
({\bf n s})n^\alpha\right]\nabla_{\bf s}^\alpha\!-\!
\nonumber\\&&\frac{(2-\gamma)^2}{\gamma^2(d-1)}
\left[-\left(\frac{1}{2}+\frac{1}{d+1-\gamma}\right)s^2+
\left(2-\gamma/2+\frac{1}{d+1-\gamma}\right)
({\bf n s})^2\right],
\label{A2a}
\end{eqnarray}
where ${\bf n}=\bbox{\rho^{cl}}/\rho^{cl}$.
It is seen clearly from (\ref{A2a}) that the transversal part of 
the effective potential is not bound from below.
Accounting for nonlinear terms is required to regularize the divergence
(see Section IV and Appendix C for an explanation of how
to avoid an explicit calculation of the terms). 
By this means,
to describe the longitudinal fluctuation one should make ${\bf s}$ be parallel
to ${\bf n}$ in (\ref{A2a}) and so deal with the reduced operator
\begin{equation}
\hat{\cal L}^{\prime\prime}=
\frac{d-1}{2-\gamma}\frac{d^2}{ds^2}+\frac{(d+2-\gamma)(d-1)}{2-\gamma}
\frac{1}{s}\frac{d}{ds}+\frac{4-3\gamma}{2\gamma}s\frac{d}{ds}-
\frac{(3-\gamma)(2-\gamma)^2}{2\gamma^2(d-1)}s^2.
\label{A2b}
\end{equation}
In the long time ($\tau\gg 1$) asymptotic the major contribution
into $\Psi_T$ stems from the lowest eigenstate of $\hat{\cal L}^{\prime\prime}$. 
The eigenfunction of the ground state is an exponential of the quadratic form
\begin{equation}
\Psi_{gr}\sim\exp\left[-\frac{a s^2}{2}\right].
\label{Gr1}
\end{equation}
In the Heisenberg representation the eigenfunction appears as 
$\exp[\Delta_E\tau]\Psi_{gr}$, where the ground state energy for
the eigenvalue problem gets
\begin{equation}
\Delta_E=\frac{2+d-\gamma}{2\gamma}\left[2-\frac{3}{2}\gamma+\sqrt{(2-3\gamma/2)^2+
2(2-\gamma)(3-\gamma)}\right].
\label{Estr}
\end{equation}

\subsection{Inner fluctuations of the points}

${\cal K}$, ${\cal A}$ and ${\cal B}$ responsible for the inner fluctuations
of a point (disk) are estimated as
\begin{equation}
{\cal K}_\delta\sim \delta\rho^{2-\gamma},\quad 
{\cal A}_\delta\sim p\delta\rho^{1-\gamma},\quad
{\cal B}_\delta\sim p^2\delta\rho^{-\gamma},
\label{KAB}
\end{equation}
where $\delta\rho(t)$ is related to $\rho(t)$ via (\ref{pp2},\ref{pp3}).
Thus, instead of (\ref{taus}) one has to perform the following transformation
to the dimensionless $\tau_\delta,{\bf s}_{\delta i}$ variables
\begin{equation}
\tau_\delta=\frac{\gamma}{2}\int_{r^\prime}^\rho\frac{d\rho}{\rho}
\left(\frac{\delta\rho}{\rho}\right)^{1-\gamma},\quad
\delta{\bf s}_{\delta i}=\frac{\sqrt{D}R^{\gamma/4}\tilde{\bf r}_i}
{\sqrt{T\delta\rho\rho^{1-\gamma/2}}},
\label{tausp}
\end{equation}
to get rid of the temporal dependence of the Hamiltonian.
The resulting Quantum mechanics 
will yield positive spectrum with a constant gap.
The dependences of $\Delta_E$ and $\delta_E$ on $n$  are estimated as
$\sim\mbox{const}$ and $\sim 1/n$ respectively
(the potential ${\cal K}$-term has an extra smallness $\sim 1/n^2$, 
while the number of elementary excitations is $\sim n$).
The dimensionless time of evolution is short 
$\sim [\delta\rho(0^+)/\rho(0^+)]^{1-\gamma}/[1-\gamma]\ll 1$,
if $\gamma$ is not too close to $1$.
These collected observations
result in the following asymptotic for $\Psi_{\delta T}$
(analog of (\ref{PT})),$\Psi_{\delta T}\to \delta\chi\left(\tilde{\bf r}
[L/r^\prime]^{1-\gamma/2}\right)$
(we have calculated (\ref{pp3}) here).
The respective contribution of the fluctuations is thus scale ($r$) independent
\begin{equation}
{\cal Z}^{fl}_{\delta}\sim
\int\prod_{\mbox{drop}}d\tilde{\bf r}_i
\frac{\exp\left[-\tilde{r}_i^\alpha
 [\hat{\cal G}^{-1}]_{ij}^{\alpha\beta}\tilde{r}_j^\beta/2
\right]}{\sqrt{|\det[\hat{\cal G}]|}}
\delta\chi(\tilde{\bf r}[L/r^\prime]^{1-\gamma/2})\sim
q^{2dn},\quad q=\mbox{max}\left[1; L^\gamma/[nDT]\right],
\label{Zflfp}
\end{equation}
where, accounting for scale dependence only, we use
$\delta\chi\left(\tilde{\bf r}[L/r^\prime]^{1-\gamma/2}\right)$, 
as a function of $\tilde{\bf r}$,
decays on the scale $r^{\prime 1-\gamma/2}L^{\gamma/2}\sim r L^\gamma/[nDT]$.

\section{Integration over the soft rotation mode}

The measure of functional integration in (\ref{4a}) is quasi-invariant
with respect to a slight rotation of the fields
\begin{equation}
\bbox{\rho}_i(t)\to \hat{U}^{-1}(t)\bbox{\rho}_i(t),\quad
 {\bf p}_i(t)\to \hat{U}^{-1}(t)\bbox{\rho}_i(t),\quad
\det[\hat{U}(t)]=1,\quad \hat{U}(T)=\hat{U}(0)=\hat{1},
\label{A1}
\end{equation}
where $\hat{U}(t)$ is a unitary matrix realizing the $d\times d$ representation
of $SU(d)$.
Quasi-invariance means that an extra term appears at the transformation of
the action (\ref{4b})
\begin{equation}
\triangle {\cal S}=
\int_0^T
\left[\hat{U}[\hat{U}^{-1}]^\prime\right]^{\alpha\beta}
\rho_i^\alpha p_i^\beta dt.
\label{A2}
\end{equation}
To fix the gauge quasi-invariance we will use a general method
\cite{67FP}, that is popularly known in field theory.
To integrate over the soft mode we will put under the functional integration
(\ref{4a}) the unity 
\begin{equation}
1=\int {\cal D}\hat{U}(t)
\delta\left(\left[\left(\hat{U}\bbox{\rho}_1\right)^\prime\times
\hat{U}\bbox{\rho}_1\right]\biggl/\rho_1^2\right)\det\left[\delta 
\left[\left(\hat{U}\bbox{\rho}_1\right)^\prime\times
\hat{U}\bbox{\rho}_1\biggl/\rho_1^2\right]
\biggl/
{\delta \hat{U}}\right],
\label{A3}
\end{equation}
where, ${\bf n}_1=\bbox{\rho}_1/\rho_1$ and $\times$ stands 
for the antisymmetric vector product of $d$ dimensional vectors.
Thus, we use a requirement for one of the particles (labeled by $1$),
to not rotate over the origin, as
a gauge condition (without loss of generality one can choose
an arbitrary direction, characterized the trajectory, to be nonrotating).
For the sake of simplicity, 
let us consider the two dimensional version of (\ref{A3})
\begin{eqnarray}&&
1=\int {\cal D}\varphi(t)
\delta\left(\left[\left(\hat{U}\bbox{\rho}_1\right)^\prime\times
\hat{U}\bbox{\rho}_1\right]\biggl/\rho_1^2\right)\det\left[\delta 
\left[\left(\hat{U}\bbox{\rho}_1\right)^\prime\times
\hat{U}\bbox{\rho}_1\biggl/\rho_1^2\right]
\biggl/
{\delta \varphi}\right],\label{A4}\\&& \hat{U}(t)=
\left(\begin{array}{cc} \cos\varphi &\sin\varphi\\
-\sin\varphi & \cos\varphi\end{array}\right).
\nonumber
\end{eqnarray}
Direct calculation gives 
\begin{equation}
\left(\hat{U}\bbox{\rho}_1\right)^\prime\times
\hat{U}\bbox{\rho}_1\biggl/\rho_1^2=\dot{\varphi}+
[\dot{\bbox{\rho}}_1\times\bbox{\rho}_1]/\rho_1^2.
\label{A5}
\end{equation}
Thus, the determinant from the rhs of (\ref{A4}) does not depend 
on the dynamical field $\bbox{\rho}_1$, and we can drop the determinant.
As a next step let us perform the change of variables (\ref{A1}) in the
original functional integral (\ref{4b}) (with the unity (\ref{A4}) substituted
into the integrand).
The Jacobian of such a transformation is unity. The $\delta$- function 
from the rhs of (\ref{A4}), describing the gauge condition, turns out to be
$\varphi$-independent. We get finally that the only factor  calculated
integration over the soft (gauge) mode is
\begin{equation}
\int{\cal D}\varphi \exp\left[-\Delta S\{\varphi\}\right],\quad
\Delta S=
\int_0^T\dot{\varphi}
\varepsilon^{\alpha\beta}
\rho_i^\alpha p_j^\beta dt,
\label{A6}
\end{equation}
where $\hat{\varepsilon}$ is the antisymmetric $2\times 2$ tensor.
Further, the integration over ${\cal D}\varphi$, being performed,
reduces (\ref{A6}) to the $\delta$-function
\begin{equation}
\delta\left(\left[ 
\varepsilon^{\alpha\beta}
\rho_i^\alpha p_i^\beta
\right]^\prime\right).
\label{A7}
\end{equation}
The condition under the $\delta$-function, which is satisfied
on a saddle-point solution as describing conservation of the angular momentum
(\ref{M}), is thus valid for fluctuations too.

%\end{multicols}

\begin{references}

\bibitem[*]{Prin} Address from September 1, 1996:
Department of Physics, Princeton University,
Princeton, NJ 08544.

 
\bibitem{68Kra-a}
R.~H. Kraichnan, 
%``Small scale structure convected by turbulence'',
Phys.~Fluids {\bf 11},  945  (1968).

\bibitem{94Kra} R.~H.~Kraichnan,
%``Anomalous Scaling of a Randomly Advected Passive Scalar'',
Phys.~Rev.~Lett. {\bf 72}, 1016 (1994).

\bibitem{95KYC}  R.~H.~Kraichnan, V.~Yakhot and S.~Chen, 
% ``Scaling relations for a randomly advected passive scalar field'', 
Phys.~Rev.~Lett. {\bf75}, 240 (1995).

\bibitem{95CFKLb}
M.~Chertkov, G.~Falkovich, I.~Kolokolov, and V.~Lebedev, Phys.~Rev.~E {\bf 52}, 
4924 (1995).

\bibitem{95GK}
K.~Gaw\c{e}dzki and A.~Kupiainen, 
%``Anomalous scaling of the passive scalar'',
Phys. Rev. Lett. {\bf 75}, 3608 (1995).

\bibitem{95CF}
M.~Chertkov and G.~Falkovich, Phys.~Rev.~Lett. {\bf 76}, 2706 (1996).

\bibitem{96BGK}
D.~Bernard, K.~Gaw\c{e}dzki and A.~Kupiainen, 
%``Anomalous scaling in the $N$-point functions of passive scalar'',
%chao-dyn/9601018, 
Phys.~Rev.~E {\bf 54}, (1996).

\bibitem{77Lip}L.~N.~Lipatov, 
%``Divergence of the perturbation theory series and the 
%quasi-classical theory'', 
Sov. Phys. JETP {\bf45}, 216 (1977).

\bibitem{94SS} 
B.~I.~Shraiman and E.~D.~Siggia, Phys. Rev. E
{\bf 49}, 2912 (1994).

\bibitem{Kra} R.~H.~Kraichnan (unpublished).


\bibitem{95CFKLa} M.~Chertkov, G.~Falkovich, I.~Kolokolov and V.~Lebedev,
Phys.~Rev.~E {\bf 51}, 3974 (1995).

\bibitem{94CGK} M.~Chertkov, A.~Gamba and I.~Kolokolov,
Phys.~Lett~A {\bf 192}, 435 (1994).

\bibitem{95BCKL} E.~Balkovsky, M.~Chertkov, I.~Kolokolov and V.~Lebedev,
JETP Lett {\bf 61}, 1012 (1995).

\bibitem{95FKLM} G.~Falkovich, I.~Kolokolov, V.~Lebedev and A.~Migdal,
``Instantons and Intermittency'',
chao-dyn/9512006, submitted to Phys.~Rev.~E.

\bibitem{76Dom}
C.~de~Dominicis, J.~Physique (Paris) {\bf 37}, c01-247 (1976).
 
\bibitem{76Jan}
H. Janssen, Z.~Phys.~B {\bf 23}, 377 (1976).

\bibitem{95GM} V.~Gurarie and A.~Migdal,
``Instantons in Burgers' Equation'',
hep-th/9512128, submitted to Phys.~Rev.~E.

\bibitem{96BFKL} E.~Balkovsky, G.~Falkovich, I.~Kolokolov, and V.~Lebedev,
``Viscous Instanton for Burgers' Turbulence'', chao-dyn/9603015.

\bibitem{95SS} B.~Shraiman and E.~Siggia, C.R. Acad. Sci. Paris, {\bf 321}, 
279 (1995).

\bibitem{95Fri} U.~Frisch, ``Turbulence. The legacy of A.N. Kolmogorov.'',
Cambridge Univ. Press, Cambridge, (1995).

\bibitem{93Maj} A.~Majda, J.~Stat.~Phys. {\bf 73}, 515 (1993).

\bibitem{94Fal} G.~Falkovich, Phys.~Rev.~E {\bf 49}, 2468 (1994).

\bibitem{95Yak} V.~Yakhot, "Passive Scalar Advected by a White-in-Time
Random Velocity Field. Probability Density of Scalar Differences",
Princeton University preprint (1995), submitted to Phys.~Rev.~E.

\bibitem{95Pol} A.~Polyakov, Phys.Rev.E {\bf 52}, 6183 (1995).

\bibitem{96SS} B.~Shraiman and E.~Siggia, ``Symmetry and Scaling of Turbulent
Mixing'', cond-mat/9604024.

\bibitem{96CFL} M.~Chertkov, G.~Falkovich and V.~Lebedev,
Phys.~Rev.~Lett. {\bf 76}, 3707 (1996).

\bibitem{84AHGA} R.~A.~Antonia, E.J. Hopfinger, 
Y.~Gagne, and F.~Anselmet, Phys.Rev.A {\bf 30}, 2704 (1984).

\bibitem{91Sre} K.~R.~Sreenivasan, Proc.~R.~Soc.~London~A {\bf 434}, 165 (1991).

\bibitem{67FP} L.~D.~Faddeev and V.~N.~Popov, Phys.~Lett.~{\bf 25B}, 29 (1967).



\end{references}
\end{document}